\documentclass[]{jfm}

\usepackage{amssymb}
\usepackage{amsmath}
\usepackage{mathrsfs}
\usepackage{layouts} 
\usepackage{printlen}
\usepackage{graphicx} 
\usepackage{natbib}
\usepackage{pgf}
\usepackage{wasysym}

\begin{document}

\newtheorem{lemma}{Lemma}
\newtheorem{corollary}{Corollary}

\shorttitle{SOS-based control of laminar wake flows} 
\shortauthor{Lasagna et al.} 

\title{Sum-of-Squares approach to feedback control of laminar wake flows}

\author
 {
 Davide Lasagna\aff{1}
  \corresp{\email{davide.lasagna@soton.ac.uk}},
 Deqing Huang \aff{2}
  ,
 Owen R. Tutty \aff{1}
  \and
 Sergei Chernyshenko\aff{3}
 }

\affiliation
{
\aff{1}
Engineering and the Environment, University of Southampton, Highfield, Southampton, SO17 1BJ, UK
\aff{2}
\textcolor{black}{Institute of Systems Science and Technology,} School of Electrical Engineering, Southwest Jiaotong University, Chengdu, 610031, China
\aff{3}
Department of Aeronautics, Imperial College London, Prince Consort Road, London SW7 2AZ, UK
}

\maketitle

\begin{abstract}
In this paper a novel nonlinear feedback control design methodology for incompressible fluid flows aiming at the optimisation of long-time averages of flow quantities is presented. It applies to reduced-order finite-dimensional models of fluid flows, expressed as a set of first-order nonlinear ordinary differential equations with the right-hand side being a polynomial function in the state variables and in the controls. The key idea, first discussed in \citet{philo}, is that the difficulties of treating and optimising long-time averages of a cost are relaxed by using the upper/lower bounds of such averages as the objective function. In this setting, control design reduces to finding a feedback controller that optimises the bound, subject to a polynomial inequality constraint involving the cost function, the nonlinear system, the controller itself and a tunable polynomial function. A numerically tractable and efficient approach to the solution of such optimisation problems, based on Sum-of-Squares techniques and semidefinite programming, is proposed.

To showcase the methodology, the mitigation of the fluctuation kinetic energy in the unsteady wake behind a circular cylinder in the laminar regime at $\Rey=100$, via controlled angular motions of the surface, is numerically investigated. A compact reduced-order model that resolves the long-term behaviour of the fluid flow and the effects of actuation, is first derived using Proper Orthogonal Decomposition and Galerkin projection. In a full-information setting, feedback controllers are then designed to reduce the long-time average of the resolved kinetic energy associated with the limit cycle. These controllers are then implemented in direct numerical simulations of the actuated flow. Control performance, total energy efficiency, and the physical control mechanisms identified are analysed in detail. Key elements of the methodology, implications, and future work are finally discussed.
\end{abstract}

\section{Introduction}
In the last decades, the coordinated efforts of laboratory experiments using high-resolution flow diagnostics and large-scale direct numerical simulations have considerably progressed our understanding of key physical processes and mechanisms in turbulent flows. Despite these new discoveries, progress in the ability to effectively control their spatio-temporal evolution in complex geometries has remained more elusive, owing to the nonlinear, multi-scale nature of turbulent motion. Interest in control is driven by the huge economic, societal and environmental benefits that advances in the field would provide. Hence, the development of active control strategies is commonly regarded as one of the key enablers for future advances in efficient transportation, energy generation and distribution, and in many other technologically-relevant industrial sectors.

Controlling and mitigating large{-}scale velocity fluctuations in the flow around bluff bodies, the problem that we discuss in this paper, is one of such instances. When the Reynolds number exceeds a critical value, the periodic generation and shedding of organised vortical structures from the body produces intense fluctuations in the aerodynamic forces, resulting in structural fatigue \citep{Sarpkaya2004} acoustic noise production \citep{Blevins1984,Thomas2008} and other undesirable effects, such as vortex-induced vibrations \citep{CHK_Williamson2004}. The technological relevance of these flows has thus spawned significant interest in devising control strategies to tame their evolution. A variety of actuation/sensing strategies and control design methods have been proposed, as recently reviewed by \citet{Choi2008}.

Because of the simplicity of the geometry, the two-dimensional flow past a circular cylinder has become the paradigmatic flow model to investigate vortex dynamics around bluff bodies. The laminar, steady solution is characterised by two recirculation eddies, whose length grows linearly with $\Rey$ \citep{Fornberg1985} and becomes unstable in a Hopf bifurcation at $\Rey\approx 47$ \citep{Provansal1987, Noack1994} due to a symmetry-breaking unstable global mode \citep{Tang1997}. The ensuing nonlinear regime saturates in a sustained periodic motion, vortex shedding, a stable periodic orbit attracting trajectories in an appropriately-defined phase space of the system \citep{Rempfer2000, noack-hierarchy} before the occurrence of other bifurcations at higher $\Rey$ \citep{Barkley1996-qp}.

Control of this specific regime became a useful benchmark problem to develop and test novel feedback control design strategies \citep{Lehmann2005}. One of the perspectives adopted in several investigations on control has been the stabilisation by feedback of the unstable, steady, laminar wake flow. At low supercritical Reynolds numbers, only one unstable global mode, the K\'arm\'an mode, exists. Hence, proportional control strategies, where the signal from a single sensor located at some point in the wake is multiplied by a constant gain and fed back to the actuator, have been considered extensively, e.g. \citet{Berger1967, Monkewitz1991, Roussopoulos1993, Park1994}. \textcolor{black}{In the light of direct numerical simulation and reduced{-}order modelling techniques for linear systems, \citet{Illingworth2014} review succinctly some of these efforts and discuss the ``gain window effect'' observed in previous numerical and experimental works, i.e. when suppression of the wake instability is achieved only if the gain is within a certain interval. They show that such an effect does not result from the destabilisation by control of other unstable modes, but it is rather driven by the properties of the closed-loop system, in particular by the time delays associated with the feedback arrangement}. The authors also showed that the window shrinks as $\Rey$ increases and it does not {exist} any more at $\Rey=80$, reflecting the objective difficulty/impossibility of obtaining stabilising controllers as the Reynolds number increases. They concluded suggesting that better control strategies, with more complicated dynamics than proportional control might be required to improve performance.

\citet{Camarri2010} proposed a linear feedback design method, for the flow past a square cylinder, based on linearised dynamics and global linear stability analysis of the equilibrium solution, inspired by previous works on passive control design methods, see \citet{Giannetti2007, Marquet2008} and references therein. Camarri and Iollo proceed by examining the sensitivity of the linear stability problem with respect to the controller parameters, in order to displace the eigenvalues of the unstable and least stable modes to the left half of the complex plane via control design. They pointed out that the performance of this controller far from the design state, i.e. the control of the nonlinear saturated regime, needs to be explored \textit{a posteriori}. They show that their feedback strategy can stabilise the fully nonlinear regime up to twice the critical Reynolds number of the natural flow. At higher Reynolds number, in highly nonlinear regimes, performance worsens. Interestingly, the authors point out that the basin of attraction of the stabilised wake structure shrinks consistently as the Reynolds number increases.

\citet{carini2015feedback} investigated feedback control in the framework of linear optimal control theory and designed and tested a full-dimensional Minimum-Control-Energy compensator, free from spill-over effect induced by the excitation by actuation of stable dynamics, often observed when control is designed on a reduced-order model \citep{Barbagallo2009}.  Using the feedback from a single sensor measuring the cross-stream velocity to control the rotation rate of the cylinder around its axis, the authors showed that complete stabilisation of the unstable mode was possible only up to $Re\approx59$, if the sensor was located in a narrow region between 2 and 2.5 diameters downstream of the cylinder axis. The critical $Re$ was increased to 72 when a full-information controller was employed. \textcolor{black}{They commented on this difference by suggesting that better performance on the nonlinear saturated flow could be obtained by adopting a nonlinear observer and ultimately a nonlinear control strategy.}

These and other investigations have demonstrated that, as the Reynolds number increases, the flow dynamics becomes so strongly nonlinear to render linearisation of the equations around the unstable equilibrium and linear design methods scarcely effective. In the terminology of \citet{Brunton2015}, such systems are ``irreducible'', in the sense that key nonlinear processes, such as vortex pairing/merging, inter-modal energy transfers, advection of coherent structures, crucial to describe the developed state of natural instabilities that arise progressively as the Reynolds number increases, cannot be described by a linearised theory. Furthermore, the gradual loss of linear stabilizability as the Reynolds number increases \citep{Lauga2003}, coupled with sensing/actuation constraints of practical technological nature, suggest that the developed nonlinear state of the flow needs to be addressed directly in the design stage.

Strategies where the structure of the feedback controller is heuristically fixed a priori and appropriate gains are obtained from optimisation or parameter exploration over nonlinear controlled regimes have been proposed, e.g. \citet{Fujisawa2001,Siegel2006,Weller2009,Lu2011,Mehmood2014}. \citet{Weller2009} introduced a feedback structure consisting of a linear proportional controller relating several cross-flow velocity measurements in the near wake to the signal driving the actuators, two blowing/suction slots on the top and bottom walls of the square cylinder arrangement driven in opposite phase. Optimisation of the gains, to reduce the short-time-averaged $L_2$-norm of the difference between the instantaneous flow field and the unstable steady solution, was then performed in a trust-region, reduced-order, adaptive setting. The resulting feedback arrangement was able to stabilise the flow starting from the saturated nonlinear regime at a Reynolds number almost twice the critical value. However, because  the optimisation involved a cost function defined over a finite, short horizon, the best controller resulted in excellent performance in this interval but performance subsequently degraded, especially at larger Reynolds numbers. The authors concluded by pointing out that the asymptotic stability of the closed-loop system cannot be ensured by their method, as the long-term behaviour of the system is not considered in the design.

These investigations strongly relied on the ingenuity of the researchers, on heuristic choices of sensor/actuation position and type, and on solid understanding of the flow physics. Such heuristic strategies might show significant limitations when applied to flows with more complex nonlinear dynamics. \textcolor{black}{Recent model-free approaches, such as genetic programming control (see e.g. \citet{Debien,Parezanovic} and reference therein), use evolutionary strategies to automatically discover such heuristics in experimental control studies, using a black-box optimisation approach. These approaches can lead to emergence of unexpected control solutions, as they effectively explore large search spaces, and can uncover novel control mechanisms \citep{Gautier}.}

\textcolor{black}{On the other side of the spectrum,} optimal control theory \citep{Abergel1990} is likely one of the most versatile model-based control design method for nonlinear systems. Optimal control, in the predictive setting, consists in finding and applying in a feed-forward fashion the control input that optimises a suitable cost function defined as an definite integral spanning a predetermined finite-horizon. Although such a strategy is extremely computationally expensive, it is considered to represent the upper limit on the achievable control performance \citep{Bewley2001}. Optimal control of a circular cylinder wake via rotary actuation has been implemented by \citet{protas2002optimal} to minimise a cost function involving the sum of the power associated with control and that associated with the drag, using optimisation horizons up to roughly one vortex shedding period. More recently, \citet{Flinois2015} implemented the same algorithm but significantly extended the optimisation horizon, up to 100 convective time units, i.e. at least ten times larger than previous efforts. The important observation is that long-time horizons, representative of the long-term behaviour of the controlled system, were necessary to suppress vortex shedding at Reynolds numbers between 75 and 200, and achieved far better performance, with smoother control inputs, than the short-time horizon approach of \citet{protas2002optimal}, enabling the identification of physical control mechanisms.

\subsection{Objectives and structure of the paper}
The main purpose of this paper is to present a novel paradigm for model-based feedback control of fluid flows, in an effort to address some of the outstanding issues discussed in the introduction.  \textcolor{black}{Firstly, the proposed control paradigm  applies directly to nonlinear Galerkin-type models of incompressible fluid flows. No linearisation around an operating point is performed and the only dynamical approximation is the truncation of the Galerkin velocity expansion. Hence, important nonlinear processes that can be described by such models} can be controlled, if not exploited. Secondly, the long-term behaviour of the system is central in the design, as the optimisation targets long-time averages, defined over infinite horizons. The key step to overcome the objective difficulty of treating and optimising such averages is to replace it by estimation/optimisation of bounds, as first proposed by \citet{philo}.

The theoretical and algorithmic back-bone of this approach is a recent breakthrough in control theory and optimisation, i.e. the discovery that the Sum-of-Squares (SOS) decomposition of a polynomial can be found, if it exists, via the solution of a semidefinite  program (SDP) \citep{parrilo2003semidefinite}. These advances have recently emerged as a promising basis to solve many computationally hard analysis/design problems for systems whose dynamics are described by polynomial functions, such as the estimation of the attraction region of equilibria \citep{Valmo2013} as well as the simultaneous optimisation of a stabilising controller and a high-degree Lyapunov function certifying the stability of the controlled system \citep{Zhao2010,Nguang2011}.  These new paradigms of design and analysis provide us with numerically tractable methods to take a new perspective of many fundamental problems in fluid dynamics as reviewed in \citet{philo}, such as nonlinear control design, the objective of this paper, or nonlinear stability analysis, as in \citet{Huang2015}.

The paper is organised as follows: in section \ref{sec:sos}, a concise presentation of Sum-of-Squares techniques is reported. Numerous references on this technology are presented for the more interested reader. Section \ref{sec:the-methodology} describes the control design methodology, using a relatively general notation. More specifically, it discusses the technique used to estimate bounds on long-time averages and its application to control design via bounds optimisation. In section \ref{sec:application} these ideas are applied to the benchmark control problem of regulation of vortex shedding past a circular cylinder at a Reynolds number equal to 100, via rotary oscillations of the surface. This problem was extensively discussed in this introduction to put our results in a more focused context and was chosen as a pretext to describe a methodology that applies independently of the specific case, i.e. from the details of the flow, the actuation/sensing arrangement and the modelling approach. The numerical setup is discussed first. The model order reduction strategy, based on Proper Orthogonal Decomposition and Galerkin projection, is then introduced. State-feedback controllers are further designed and performance is assessed by implementation in direct numerical simulation (DNS) in a full-information setting. Conclusions and future work to be addressed are summarised in section \ref{sec:conclusions}.

\section{The Sum-of-Squares decomposition}\label{sec:sos}
We provide in this section a succinct overview of Sum-of-Squares techniques, in order to convey the general underlying ideas. In this section, we favour clarity over mathematical rigour, with the hope of bridging the gap between the mathematical aspects and fluid mechanics. We refer the interested reader to our previous work \textcolor{black}{\citep{philo}}, where a broader overview of Sum-of-Squares techniques and its applications in fluid mechanics is given.

Despite the complexity of the underlying mathematical framework, the idea of the Sum-of-Squares (SOS) decomposition of a polynomial is rather simple. As an example, one might be interested in checking the non-negativity of a given multivariate polynomial function $P(a_1, \ldots, a_N) = P(\mathbf{a})$, of even degree $2d_P$, that is if $P(\mathbf{a}) \ge 0~\forall \mathbf{a} \in \mathbb{R}^N$. Checking non-negativity for a general multivariate polynomial is NP-hard, hence intractable from a computational perspective \citep{papachristodoulou2002construction}. However, a sufficient condition for $P(\mathbf{a})$ to be non-negative is that it can be decomposed into the sum of the squares of $M$ polynomial functions $p_1(\mathbf{a}), \ldots, p_M(\mathbf{a})$, of lower \textcolor{black}{degrees} as
\begin{equation}
	P(\mathbf{a}) = \sum_{i=1}^M p_i^2(\mathbf{a})
\end{equation}
Finding such a decomposition is equivalent to finding a positive semidefinite matrix $\mathbf{Q}$, which can be assumed symmetric without loss of generality,  and a suitable vector $\mathbf{v}(\mathbf{a})$ containing monomials in \textcolor{black}{the variables $a_i$ up to degree $d_P$ such that}
\begin{equation}\label{eq:sos-quadratic-form}
	P(\mathbf{a}) = \mathbf{v}(\mathbf{a})^T  \mathbf{Q} \mathbf{v}(\mathbf{a})
\end{equation}
If one can find a positive semidefinite $\mathbf{Q}$, then a linear transformation of coordinates can reduce it to a diagonal form, with non-negative entries on the main diagonal, reducing $P$ to a linear combination of squares of polynomials, clearly being equivalent to non-negativity. However, the converse is not necessarily true, that is not all non-negative polynomials admit a SOS decomposition, a famous counter example being the Motzkin polynomial.

In a design problem, it might be of interest to \emph{construct} a non-negative polynomial function subject to a set of constraints, rather the checking non-negativity of an existing one. This problem, which we will deal with in what follows, can be treated essentially using the same approach. It is worth noticing that, in practice, the decomposition (\ref{eq:sos-quadratic-form}) is only approximate, and the error
\begin{equation}\label{eq:sos-quadratic-form1}
	e(\mathbf{a})=P(\mathbf{a})-\mathbf{v}(\mathbf{a})^T  \mathbf{Q} \mathbf{v}(\mathbf{a})
\end{equation}
is nonzero. However, by Theorem 4 in \citet{Lofberg2009-mu}, the polynomial $P(\mathbf{a})$ is still certifiably non-negative if
\begin{equation}\label{post-check}
	\lambda_{min}-\dim(\mathbf{Q})\times |r|\ge 0,
\end{equation}
where $\lambda_{min}$ is the smallest eigenvalue of $\mathbf{Q}$, \textcolor{black}{$\dim(\mathbf{Q})$ denotes the dimension of the matrix $\mathbf{Q}$,} and $r$ be the coefficient of $e(\mathbf{a})$ that has the largest magnitude.

From a computational perspective, finding the SOS decomposition of a polynomial amounts to finding a positive semidefinite matrix $\mathbf{Q}$, subject to a set of linear equality constraints, arising from the equality in (\ref{eq:sos-quadratic-form}). This problem can be reformulated as a semidefinite programme (SDP) \citep{parrilo2003semidefinite}, a convex and tractable problem to solve. Several freely-available software tools that can formulate and solve efficiently this kind of problems exists, such as the Matlab toolboxes SOSTOOLS \citep{sostools}, and YALMIP \citep{YALMIP}.

\section{The control design method}\label{sec:the-methodology}
\subsection{Problem statement}
We consider finite-dimensional dynamical systems given as a set of nonlinear, coupled ordinary differential equations, as
\begin{equation}\label{eq:sys}
	\frac{\mathrm{d}\mathbf{a}}{\mathrm{d}t}=\mathbf{f}(\mathbf{a}, u)
\end{equation}
where $\mathbf{a} \in \mathbb{R}^N$ is the state variables vector, $u \in \mathbb{R}$ is the control, and $\mathbf{f} : \mathbb{R}^N \times \mathbb{R} \rightarrow \mathbb{R}^N$ is assumed to be a \emph{polynomial} function in the state variables and in the control. For the sake of reducing clutter, we discuss a single input case, but the multiple input can be treated with minor revisions in the derivation. For incompressible fluid flows this is the formulation that results naturally from Galerkin projection of the governing equations onto a finite-dimensional orthonormal set of basis functions \citep{Fletcher1984-bj}. It is well known that, for such systems, the vector field $\mathbf{f}$ \textcolor{black}{can have constant,} linear and quadratic terms, and the latter conserves energy for a large class of boundary conditions of the original partial differential equations. The way the control appears in $\mathbf{f}$ depends on the type of actuation: for volume forces the right hand side is affine with the input $u$, whereas for actuation via the boundary a ``lifting'' procedure results in the dynamics of the system being dependent on $\mathrm{d}u/\mathrm{d}t$ and $u^2$.

Suppose that for system (\ref{eq:sys}) it is of interest to control the value of a turbulent quantity $\Phi(t)$, the cost. This could express, for instance, the instantaneous turbulent kinetic energy, or the energy dissipation rate. Suppose further that the cost can be expressed as a positive-definite polynomial function of the state variables and of the control, i.e. $\Phi(t) = \Phi(\mathbf{a}(t), u(t))$. In general, but more specifically for systems exhibiting turbulent behaviour, long-time statistics of $\Phi(t)$, for example long-time averages,
\begin{equation}
	\overline{\Phi} = \lim_{T\rightarrow \infty} \frac{1}{T}\int_0^T \Phi(\mathbf{a}(t), u(t)) \mathrm{d}t,
\end{equation}
are of primary interest, where $\mathbf{a}(t)$ is the solution of (\ref{eq:sys}), with $u=u(t)$ and for some initial condition. Denoting first by $\overline{\Phi}^0$ the long-time averaged cost without control, the objective is to design a state-feedback controller
\begin{equation}\label{eq:controller}
	u(t) = g(\mathbf{a}(t))
\end{equation}
that manipulates the long-term behaviour of (\ref{eq:sys}) such as to reduce, or increase depending on the problem, the long-time averaged cost to $\overline{\Phi}^*$. Here, we also restrict $g:\mathbb{R}^N \rightarrow \mathbb{R}$ to be an initially undetermined polynomial function of arbitrary degree $d_g$ in the state variables, in order to leverage SOS techniques. Such a restriction imposes a high degree of smoothness on the control, but highly nonlinear controllers can be designed, as $d_g$ can be regarded as a design parameter. Note that the controller (\ref{eq:controller}) makes the closed-loop system (\ref{eq:sys}) an autonomous system. We assume here that complete and exact information on the instantaneous state of the system is available; hence, we avoid the necessity of designing an observer. This step would be required in a practical application, but it is out of the scope of this paper, which focuses on control design only.
\begin{figure}
	\centering
	\includegraphics[width=0.85\textwidth]{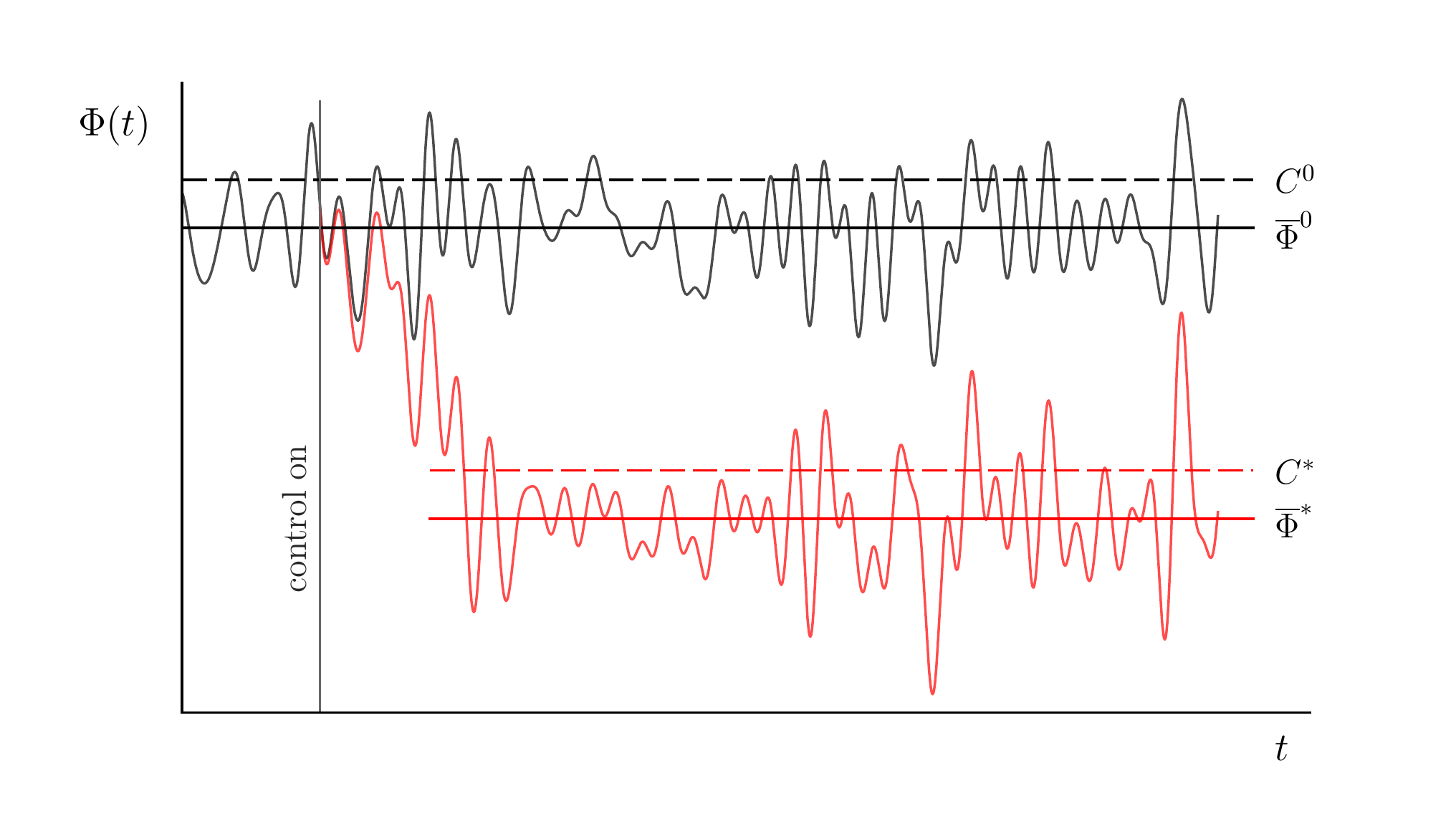}
	\caption{(Colour online). Illustration of the general idea behind the proposed control methodology. Instead of designing a controller that reduces the time average from $\overline{\Phi}^0$ to $\overline{\Phi}^*$, a controller that reduces the upper bound from $C^0$ to $C^*$ is sought. Under the action of such a controller the time average is also expected to decrease, although this cannot be guaranteed in a general case.}
	\label{fig:bounds}
\end{figure}

Ideally, the controller \textcolor{black}{could be designed} by solving the optimisation problem
\begin{equation}\label{eq:optim-average}
\overline{\Phi}^* = \left \{
\begin{array}{lll}
\underset{g}{\text{min}} &  & \overline{\Phi} \\
\text{subject to}        &  & \dot{\mathbf{a}}=\mathbf{f}(\mathbf{a}, g(\mathbf{a}))
\end{array}
\right.
\end{equation}
where $\dot{\mathbf{a}} = \mathrm{d}\mathbf{a}/\mathrm{d}t$. The non-convexity of (\ref{eq:optim-average}), but most importantly the fact that the minimisation of long-time averages are considered, \textcolor{black}{\textcolor{black}{makes} (\ref{eq:optim-average}) difficult to solve.}
The key step, previously suggested in \citet{philo}, is illustrated in figure \ref{fig:bounds}. Instead of treating a long-time average directly, \textcolor{black}{we replace the original problem} with the analysis of an upper bound of the average, i.e. a value $C$ for which an algorithm exists proving $\overline{\Phi} \le C$ for system (\ref{eq:sys}), where the equality holds when the bound is tight. Hence, instead of attempting to reduce the long-time average, we reformulate (\ref{eq:optim-average}) into the problem of designing a controller minimising the upper bound, from $C^0$, the bound on $\overline{\Phi}^0$, to $C^*$, the bound on $\overline{\Phi}^*$. This reads as
\begin{equation}
C^* = \left \{
\begin{array}{lll}
\underset{g}{\text{min}} &  & C \\
\text{subject to}        &  & \bar{\Phi}\le C, ~~\dot{\mathbf{a}}=\mathbf{f}(\mathbf{a}, g(\mathbf{a}))
\end{array}
\right.
\end{equation}
The hope is that under the action of such a controller, the actual time-average $\overline{\Phi}^*$ will also decrease. This is not guaranteed to happen in a general case. As a trivial, yet representative, example, consider a system having multiple stable equilibria $\mathbf{a}_i$, each with its own basin of attraction. In such a case, the long-term behaviour of trajectories, hence time average, depends on the particular choice of the initial conditions. An upper bound on the time average of some cost $\Phi(\mathbf{a})$ is simply $\max_i \Phi(\mathbf{a}_i)$. The crucial point is that a controller designed to reduce the upper bound is guaranteed to decrease the actual time average only in a ``worst-case scenario'', i.e. when the trajectory starts from the basin of attraction of the steady solution associated with the bound. Should this not be the case, it is perfectly possible that the actual long-time average will increase.

The occurrence of such a behaviour depends on the particular choice of the cost function $\Phi(\mathbf{a})$ and on topology of the system's phase space, i.e. the attractors/repellors that populate it. Nevertheless, manipulating and analysing bounds  is much easier than doing so with long-time averages directly. SOS techniques can be employed exactly for such a purpose as we show in the next section. In the case when the algorithm used to calculate the upper bound does not guarantee that the bound is tight the outcome of the optimisation depends also on the algorithm, which, therefore, should be specified. This is done in the following section.

\subsection{Bounds estimation}\label{sec:bound-estimation}
The first step is to derive an upper bound $C^0$ on the average $\overline{\Phi}^0$, for uncontrolled dynamics. \textcolor{black}{We define a polynomial function} in the state variables, $V(\mathbf{a})$, of degree $d_V$, containing unknown decision variables as its coefficients. \textcolor{black}{We assume that trajectories of the system (\ref{eq:sys}) are bounded in some set, as one would expect in a dissipative system such as a fluid flow, both in the infinite dimensional case, and for non-degenerate finite-dimensional representations thereof.}

The function $V$ is also bounded in this set as it is a polynomial. The total time derivative of $V$ along trajectories of the system,
\begin{equation}
    \frac{\mathrm{d}V(\mathbf{a})}{\mathrm{d}t} = \frac{\partial V}{\partial\mathbf{a}} \cdot \frac{\mathrm{d}\mathbf{a}}{\mathrm{d}t} = \nabla_{\mathbf{a}} V(\mathbf{a})\cdot \mathbf{f}(\mathbf{a}),
\end{equation}
is then also bounded, where $\nabla_{\mathbf{a}} V \triangleq \partial V/ \partial \mathbf{a}$ is the gradient of $V$ with respect to the coordinates of the phase space. Now, suppose one can find some $V$ such that the following polynomial inequality
\begin{equation}\label{eq:polynomial-inequality}
	\nabla_{\mathbf{a}} V(\mathbf{a}) \cdot \mathbf{f}(\mathbf{a}) + \Phi(\mathbf{a}) \leq C
\end{equation}
is satisfied for all $\mathbf{a}$ in $\mathbb{R}^N$, for a given $C$. Then, it is straightforward to show that $C$ is an upper bound for $\overline{\Phi}^0$, i.e. $\overline{\Phi}^0 \leq C$. This is because when taking the time average of (\ref{eq:polynomial-inequality}) with $\mathbf{a}=\mathbf{a}(t),$
 the term
\begin{equation}
	\displaystyle \overline{\nabla_{\mathbf{a}} V(\mathbf{a}) \cdot \mathbf{f}(\mathbf{a})} = \overline{\frac{\mathrm{d}V(\mathbf{a})}{\mathrm{d}t}} = \lim_{T \rightarrow \infty} \frac{1}{T}\int_0^T \frac{\mathrm{d}V(\mathbf{a})}{\mathrm{d}t} \mathrm{d}t = \lim_{T \rightarrow \infty} \frac{1}{T} [V(\mathbf{a}(T)) - V(\mathbf{a}(0))]
\end{equation}
vanishes identically under the above assumption of boundedness.
Hence, the upper bound $C^0$ can be obtained by minimizing $C$ over all possible polynomials $V$ of a given degree under the polynomial constraint (\ref{eq:polynomial-inequality}), i.e. by solving
\begin{equation}\label{eq:mini-no-control-pos}
\overline{\Phi}^0 \leq C^0 = \left \{
\begin{array}{lll}
\underset{V}{\text{min}} &  & C \\
\text{subject to}        &  & -\Big ( \nabla_{\mathbf{a}} V(\mathbf{a}) \cdot \mathbf{f}(\mathbf{a}) + \Phi(\mathbf{a}) - C \Big ) \textcolor{black}{\ge} 0
\end{array}
\right.
\end{equation}

Because verifying positive-definiteness of a given polynomial, as well as constructing one as in the present case, is a notoriously difficult problem, we replace the constraint in (\ref{eq:mini-no-control-pos}) such as to have
\begin{equation}\label{eq:mini-no-control-sos}
\overline{\Phi}^0 \leq C^0 \leq C_{SOS}^0 = \left \{
\begin{array}{lll}
\underset{V}{\text{min}} &  & C \\
\text{subject to}        &  & -\Big (\nabla_{\mathbf{a}} V(\mathbf{a}) \cdot \mathbf{f}(\mathbf{a}) + \Phi(\mathbf{a}) - C \Big ) \in \Sigma
\end{array}
\right.
\end{equation}
where $\Sigma$ is the set of all polynomials that have a sum-of-squares decomposition. From a numerical point of view, this optimisation problem is solved by trial and error by reducing $C$ until a $V$ satisfying the polynomial inequality cannot be found. For a given $C$, the search for the function $V$ is numerically reformulated into a semidefinite programme using standard software tools \citep{sostools,YALMIP}. It is a convex problem, hence can be solved efficiently and the solution, \textcolor{black}{if it exists,} is unique. In general, a hierarchy of bounds can be obtained by increasing the degree of the polynomial function $V$. Note that the same procedure can be used to estimate a lower bound, when maximisation of the time-average is of interest, by reversing the sign of the inequality in (\ref{eq:polynomial-inequality}), and change (\ref{eq:mini-no-control-sos}) to a maximisation problem.

Strengthening the \textcolor{black}{non-negativity} constraint to a SOS constraint adds conservativeness in the optimisation, in the sense that the upper bound found from (\ref{eq:mini-no-control-sos}) can be, in principle, lower than the bound that could be found if one was able to solve (\ref{eq:mini-no-control-pos}) directly, so the tightness of the obtained bound may not be guaranteed. This is because not all positive-definite polynomials can be decomposed into the sum of squares of other polynomials, although this appears to be a special case \citep{Tan2006}. However, the second problem is numerically tractable, whereas the former is not.

It is worth pointing out that finding a finite upper bound of a long-time average on a positive-definite cost, with the method defined above, does not automatically prove the boundedness of the trajectory of the system. Bounds on long-time averages are determined by the invariant sets that populate the phase space of the system, no matter what is their stability, because they define the permanent regime. The bound could be given by an unstable set, e.g. a repellor, and in the absence of other information, it is not possible to assert boundedness of the trajectories. A modification of the polynomial inequality (\ref{eq:mini-no-control-pos}), based on the idea of adding stochastic noise to (\ref{eq:sys}), to include only stable invariant sets is discussed in \citet{philo}. However, with this modification, a finite upper bound only proves boundedness of trajectories for almost all initial conditions. In fact, there might still exist an unbounded unstable set, that might allow a trajectory with a particular initial condition, on this set, to escape to infinity. The inability to find an upper bound does not prove that trajectories are unbounded. This is because the SOS constraint in (\ref{eq:mini-no-control-sos}) is stronger than the \textcolor{black}{non-negativity}  constraint in (\ref{eq:mini-no-control-pos}), resulting in $C^0 \leq C_{SOS}^0$. Hence, one could probably formulate a corner-case problem where a finite $C_{SOS}^0$ cannot be found, whereas $C^0$ exists and it is finite.

A recent discussion on such issues is given by \citet{schlegel2015long}. These authors proposed a computational procedure that can be used to prove boundedness of the trajectories. It is based on the idea of finding a globally attracting ``trapping region'', i.e. a closed set in the phase space such that all trajectories converge to this region and remain inside it once they have entered it. Their procedure is based on finding an appropriate shift in the phase space, such that the perturbation energy in this translated reference frame possesses the mathematical properties of a Lyapunov function for large deviations from the shifted origin. From a computational perspective, they employed a simulated annealing algorithm, or ad-hoc searches along particular directions in the phase space to identify the appropriate shift. However, this algorithm can only prove the existence of a trapping region, it cannot disprove it. In appendix \ref{sec:trapping region} we report an alternative and rigorous SOS-based procedure that can be used to prove the existence of a monotonically trapping region.

\subsection{Bounds optimisation}
As for the bound estimation problem, we consider a tunable polynomial function $V(\mathbf{a})$, and assume initially that trajectories remain bounded under closed-loop control. The optimisation problem equivalent to (\ref{eq:mini-no-control-sos}) is now
\begin{equation}\label{eq:mini-control-sos}
C^* \leq C_{ SOS}^* = \left \{
\begin{array}{lll}
\underset{V, g}{\text{min}} &  & C \\
\text{subject to}        &  & -\big ( \nabla_{\mathbf{a}} V(\mathbf{a}) \cdot \mathbf{f}(\mathbf{a}, g(\mathbf{a})) + \Phi(\mathbf{a}) - C \big ) \in \Sigma
\end{array}
\right.
\end{equation}
where the minimisation of the upper bound is now performed over all \textcolor{black}{possible polynomial functions} $V(\mathbf{a})$ and state-feedback polynomial controllers $g(\mathbf{a})$, of given degree $d_V$ and $d_g$, respectively. The additional degrees of freedom associated with $g$ can allow, unless one is dealing with certain pathological cases, a further reduction of the upper bound, that is $C_{SOS}^* < C_{SOS}^0$.  As previously anticipated, it is not guaranteed that the feedback controller obtained using this procedure will reduce the actual value of the time average of the cost in closed-loop simulation of the system, that is the inequality $\overline{\Phi}^* < \overline{\Phi}^0$ cannot be guaranteed to hold.

The bounds $C^0$ and $C^*$ solely depend on the analytic definition of the vector field $\mathbf{f}$, hence on the structure of the system's phase space.  Because the system's invariant sets determine the long-term evolution, hence the bound, one can see this design scheme as finding the vector field induced by $g(\mathbf{a})$ that moves/reshapes the set associated to the bound such as to reduce favourably the long-time averaged cost.

The bound optimisation problem is still non-convex, because one needs to optimise simultaneously the tunable function $V(\mathbf{a})$ and the controller $g(\mathbf{a})$,  and so the tuning variables in $V$ are multiplied by those in $g$. This problem is not directly reducible to a semidefinite programme and convex optimisation techniques cannot be readily applied. This is a well-known problem in the SOS community, and is similar to that encountered when optimising simultaneously a \textcolor{black}{globally} stabilizing feedback controller and a polynomial Lyapunov function certifying the global stability of an equilibrium \citep{Zhao2010,Nguang2011}.  Alternative iterative algorithms need to be used (see e.g. \citet{henrion2005positive} for an overview).

In this paper we have developed a similar iterative algorithm, which is described in detail in appendix \ref{sec:iterative-algorithm} and used to solve (\ref{eq:mini-control-sos}).
\textcolor{black}{
The details are as follow: 
(1) for a given upper bound $C$ and an initial controller $g(\mathbf{a})$, whose derivative can be calculated in a certain way, e.g., (\ref{eq:trick}) when $g(\mathbf{a})$ is linear, we check the feasibility of the resultant SOS problem, namely, (\ref{post-check}), by tuning $V(\mathbf{a})$. Here, the feasibility of SOS optimisation implies that the controller $g(\mathbf{a})$ is effective in reducing the upper bound to $C$;
(2) fix the optimised $V(\mathbf{a})$ and still keep $\mathrm{d}g/\mathrm{d}t$ as in (1), we further minimize the upper bound $C$ by solving the resulting SDPs in the decision variables of $g(\mathbf{a})$. Note that the Schur complement technique will be adopted to resolve the nonlinearity of the SOS problem, as demonstrated in (\ref{SOS2}). In addition, only a reduction of $\delta C$ is considered at each step; 
(3) update $\mathrm{d}g/\mathrm{d}t$ using the optimised $g(\mathbf{a})$ in (2) and repeat the procedure (1)-(2) until $C$ cannot be decreased any more. 
(4) output the optimised $C$ and the corresponding controller $g(\mathbf{a})$.
}

The non-convexity implies that it is not guaranteed that these iterations will arrive at the global minimum of the bound. Our experience suggests this is indeed the case and the optimum will typically depend on the initial guess.

\textcolor{black}{It is worth mentioning that using this bound-optimisation procedure, globally stabilizing controllers can be also designed. In particular, an SOS globally stabilizing controller for an equilibrium point $\mathbf{a}_0$ is that for which the upper bound optimisation (\ref{eq:mini-control-sos}) has solution \mbox{$C_{SOS}^*= \Phi(\mathbf{a}_0)$}, provided that $\Phi$ reaches the global minimum on this point. In other words, if the bound can be made tight to $\Phi(\mathbf{a}_0)$ via control design, then all trajectories of the controlled system must eventually converge to $\mathbf{a}_0$, to make the long-time average equal to the bound. Note that here $V$ does not posses the mathematical properties of a Lyapunov function as in the Lyapunov-based method discussed in the previous paragraphs. It is also worth saying that the optimisation stops when the bound cannot be further reduced, resulting in a controller that can still modify favourably the dynamics. Possible reasons for a premature stop include conservativeness of the SOS constraint or a degree of $V$ or $g$ lower than necessary.}



\section{Application to a fluid flow}\label{sec:application}
In this section the control design methodology described in section \ref{sec:the-methodology} is applied to a fluid flow. The mitigation of fully-developed vortex shedding, i.e. the nonlinear dynamics of the two-dimensional unsteady wake flow past a circular cylinder at low Reynolds number, $\Rey=100$, using a controlled rotary motion of the cylinder, has been selected.

\subsection{Numerical setup}
\begin{figure}
	\centering
	\includegraphics[width=0.8\textwidth]{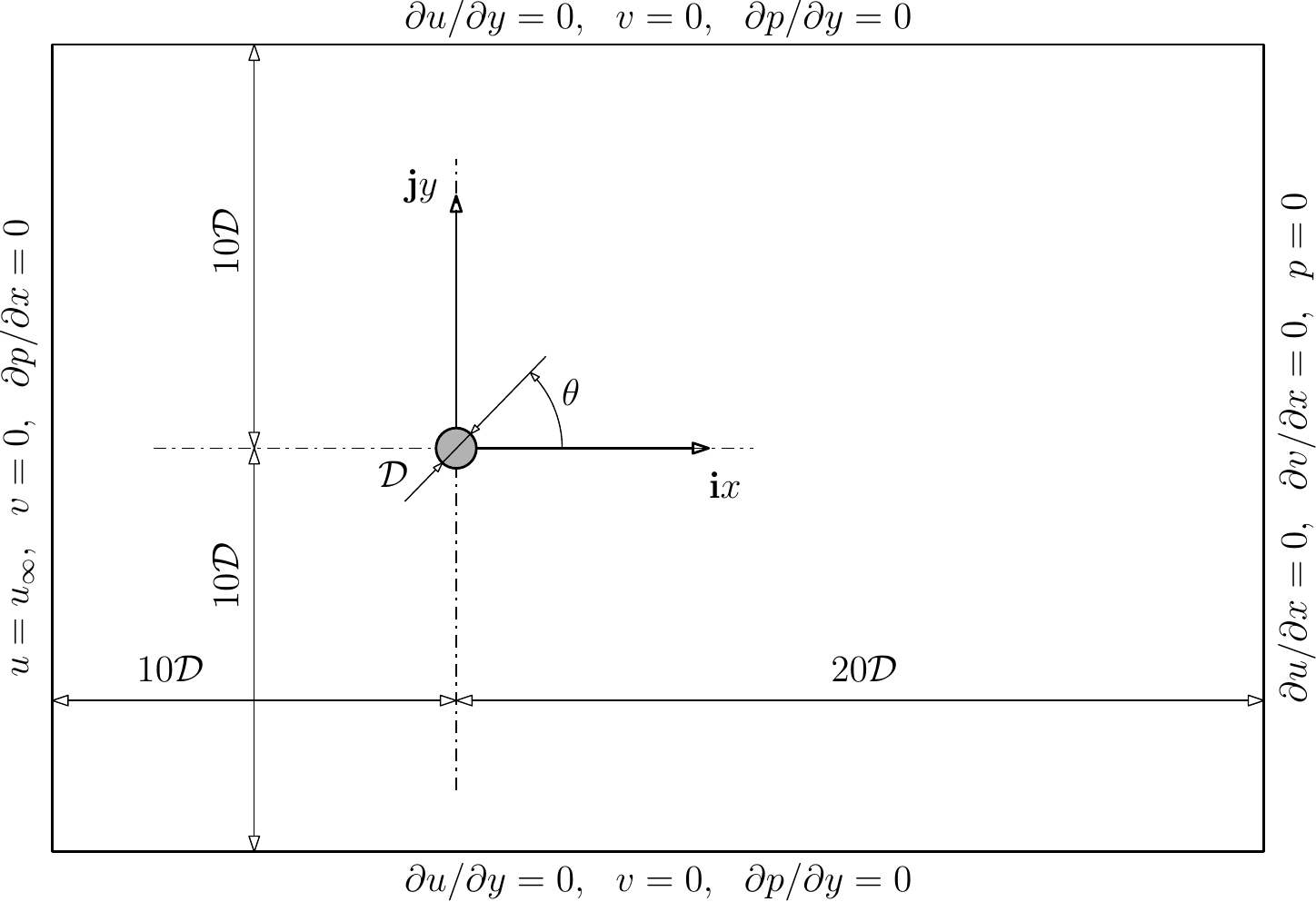}
	\caption{Schematic of the problem configuration for the circular cylinder flow. Boundary conditions on the outer domain boundaries are also indicated.}
	\label{fig:cartoon}
\end{figure}

The formulation used to solve the flow problem is based on the Navier-Stokes momentum and continuity equations for a two-dimensional incompressible viscous fluid
\begin{subequations}\label{eq:ns}
\begin{align}
	\frac{\partial \mathbf{u}}{\partial t} +\mathbf{u} \cdot \nabla \mathbf{u} &= - \nabla p + \frac{1}{Re}\nabla ^2 \mathbf{u} \\
	\nabla \cdot \mathbf{u} &= 0
\end{align}
\end{subequations}
where $p$ is the reduced pressure and $\mathbf{u} = u \mathbf{i} + v \mathbf{j}$ is the velocity vector defined on a two-dimensional Cartesian space $\mathbf{x} = x \mathbf{i} + y \mathbf{j}$, centred on the centre of the cylinder, located at $\mathbf{x} = (0, 0)$, and oriented such that the $x$ axis is aligned with the free stream, as sketched in  figure \ref{fig:cartoon}. Normalisation of the governing equations, resulting in (\ref{eq:ns}), is done using the cylinder diameter and the free stream velocity. This yields the standard definition of the Reynolds number as $Re=u_\infty \mathcal{D} /\nu$, where $\mathcal{D}$ is the cylinder diameter, $u_\infty$ is the free stream velocity and $\nu$ is the kinematic viscosity of the fluid.

The Navier-Stokes problem (\ref{eq:ns}) is solved on a triangular unstructured mesh with a finite volume formulation provided by the open source software OpenFOAM \citep{openfoam}. The application \texttt{icofoam}, implementing the well-known PISO algorithm has been used to solve the velocity-pressure coupling \citep{peric}. Preliminary validation and grid convergence studies, not reported in this paper because of the standard problem type, have been conducted to assess the reliability of the solver, showing good agreement between present and previous numerical results. A mesh of intermediate fineness with size of the elements adjacent to the cylinder equal to 0.02 has been chosen, for a total of about 17000 triangular cells.

The computational domain has the same size as the one used in \citet{bergmann2005optimal}. It is rectangular and extends for 10 and 20 diameters upstream and downstream of the cylinder, respectively, and spans a total vertical size of 20 diameters in the crossflow. The boundary conditions associated with the problem are also sketched in figure \ref{fig:cartoon}. At the inflow, the Dirichlet condition $\mathbf{u} = (u_\infty, 0)$ is used for the velocity, while the Neumann condition $\partial p/\partial x = 0$ is used for the pressure. On the upper and bottom boundaries a free-slip condition is used for the velocity, such that $\partial u/\partial y = 0$ and $v=0$. A zero-normal-gradient condition is used for the pressure on these two boundaries. On the cylinder surface the no-slip condition $\mathbf{u} = (0, 0)$ is enforced, while the standard zero normal pressure gradient condition is used for the pressure. At the outflow boundary, good numerical results, without spurious reflections, were obtained by using a zero-normal-gradient condition for the velocity, i.e. $\partial \mathbf{u}/\partial x = (0, 0)$, while the Dirichlet condition $p=0$ was set to fix uniquely the pressure field.

The time step was constant and equal to $\Delta t = 0.005$ for the mesh used to obtain all the results reported in the rest of the paper. This choice was adopted to achieve \textcolor{black}{satisfactory temporal resolution and }a maximum Courant number in the flow field on the order of 0.7.

In the following we will make use of a standard inner product between vector fields, defined as
\begin{equation}
	\langle \mathbf{v}, \mathbf{w} \rangle = \int_\Omega \mathbf{v}\cdot \mathbf{w}\,\mathrm{d}\Omega
\end{equation}
where $\Omega$ is the flow domain, and the associated norm $\|\mathbf{v}\| = \langle \mathbf{v}, \mathbf{v} \rangle^{1/2}$.

\subsection{Proper Orthogonal Decomposition and reduced-order modelling}
The SOS-based design methodology requires a finite-dimensional representation of the system dynamics, available explicitly as a set of first-order ordinary differential equations with right-hand side being a polynomial function of the state variables and of the control. Spatial discretisation of equation (\ref{eq:ns}), an infinite dimensional system modelled by partial differential equations (PDEs), leads formally to such a  system, but the extremely large dimension leads to problems that are not tractable numerically, even for moderately complex flows. Specifically, the computational cost of the solution of the SOS problems discussed above, e.g. equation (\ref{eq:mini-control-sos}), grows extremely quickly with the system size, as it will be discussed later. Hence, a model reduction strategy is used in this paper to reduce the size of the dynamical system and allow a numerically tractable solution.

We adopt a standard Galerkin projection method, whereby the full dynamics are projected onto a low-dimensional linear subspace, spanned by appropriately selected basis functions. To begin with, the velocity vector field is assumed to be approximated by the ansatz
\begin{equation}\label{eq:ansatz}
	\mathbf{u}^N(\mathbf{x}, t) = \overline{\mathbf{u}}(\mathbf{x}) + \gamma(t)\mathbf{u}(\mathbf{x}) + \sum_{i=1}^N a_i(t)\mathbf{u}_i(\mathbf{x}).
\end{equation}
The velocity field is decomposed into a solenoidal steady base flow $\overline{\mathbf{u}}(\mathbf{x})$ satisfying homogeneous boundary conditions on the cylinder, a ``control flow'' $\gamma(t)\mathbf{u}(\mathbf{x})$, \textcolor{black}{(see e.g. \citet{graham1999optimal,Kasnakoglu2008})} used to lift the time-dependent inhomogeneous boundary conditions on the oscillating cylinder surface and to include control via the boundary in the dynamic model, and the weighted sum of $N$ solenoidal vector fields $\mathbf{u}_i(\mathbf{x})$, the basis functions, which are assumed to form an orthonormal set.

Because the dynamics of a high-dimensional system are compressed into few degrees of freedom, the choice of the basis functions $\mathbf{u}_i$ is often crucial. Growing interest in model-based control of fluid flow has resulted in different selection strategies, that are far too numerous to discuss here (see, e.g., \citet{noack-hierarchy, Barbagallo2009}). In this work we used Proper Orthogonal Decomposition, POD, \textcolor{black}{\citep{Sirovich1987, berkooz, Holmes1998}} to identify the low-dimensional subspace. The motivating observation for the choice of POD in the current context is that when\ data used in the POD algorithm is specifically obtained by sampling the system after the developed regime has established, i.e. any transient has died out, the basis functions describe approximately the axis of inertia of the attractor of the system. \textcolor{black}{Hence, a ROM that describes accurately the long-term behaviour of the system is extremely important in the present case, because the focus of the present SOS paradigm in on the estimation and optimisation of a bound for the developed regime.}

\begin{figure}
	\centering
	\includegraphics[width=0.95\textwidth]{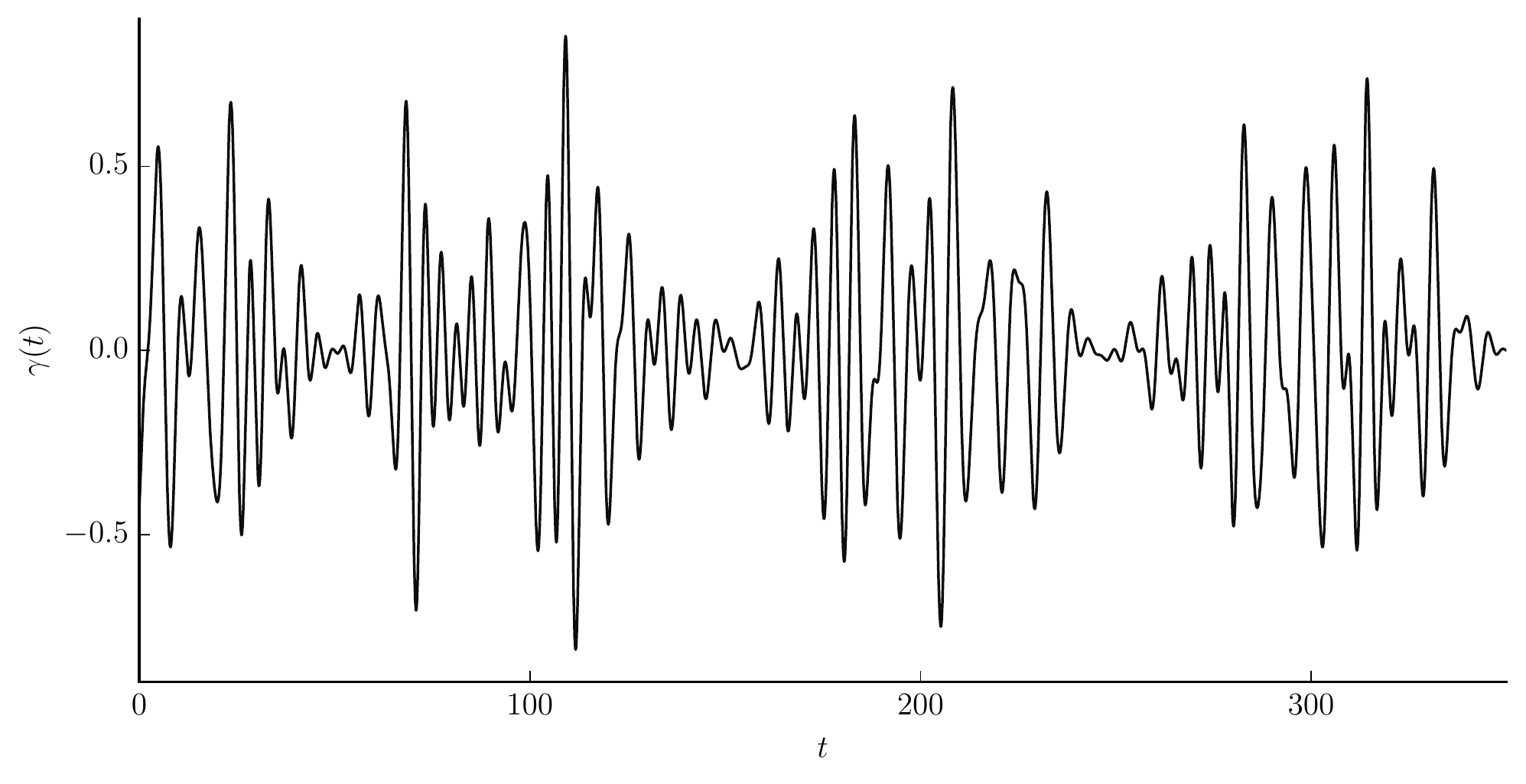}
	\caption{Time history of the signal used to generate snapshots of the actuated velocity field}
	\label{fig:control-signal}
\end{figure}

With the idea of exciting transient flow structures and obtaining a richer snapshot set, \citep{bergmann2005optimal}, a first set of snapshots of the velocity vector field, \mbox{$\mathcal{U} = \{\mathbf{u}(\mathbf{x}, t_k)\}_{k=1}^M$}, is sampled from a direct numerical simulation in which the angular motion of the cylinder is driven by a random actuation signal. The discrete-time actuation signal is obtained from samples of a zero-mean Gaussian distribution. It is then digitally filtered, such that its power spectrum has zero energy outside the band of reduced frequency $S_t = f\mathcal{D}/u_\infty = [0.1, 0.25]$.  The amplitude of the filtered signal is then modulated by a low frequency mode, $St_{mod} = 0.005$, in order to actuate the flow at different intensities, and it is then normalised to have unitary maximum magnitude, resulting in a standard deviation equal to about 0.25. One third of the total control signal is shown in figure \ref{fig:control-signal}.
The total duration of the simulation is $T=1000$, about 150 oscillation cycles of the uncontrolled flow, and a total of $M=900$ snapshots is sampled, from $t \ge 100$, at intervals of 1 nondimensional time units. We have verified that such number of snapshots is sufficient to provide convergence of the second-order statistics associated with POD, such as the energy distribution among individual POD modes.

The time-dependent, inhomogeneous boundary conditions on the cylinder are lifted from the snapshots by subtracting, with appropriate amplitude, the control function, obtaining the set
\begin{equation}
	\displaystyle \mathcal{U}' = \{\mathbf{u}^h(\mathbf{x}, t_k) = \mathbf{u}(\mathbf{x}, t_k) - \gamma(t_k)\mathbf{u}_c(\mathbf{x})\}_{k=1}^M.
\end{equation}
A radially-symmetric, \textcolor{black}{solenoidal} control function $\mathbf{u}_c(\mathbf{x})$, with circumferential velocity  decaying as $e^{-\lambda(r-0.5)}$, was used. The decay factor $\lambda=8$ was selected from a numerical simulation where the cylinder is oscillated harmonically, in quiescent fluid, with frequency equal to the shedding frequency, and monitoring the decay with $r$ of the amplitude of the velocity fluctuations in the circumferential Stokes flow.

The arithmetic average
\begin{equation}\label{eq:arithmetic-average}
	\overline{\mathbf{u}}(\mathbf{x}) = \frac{1}{M}\sum_{k=1}^M \mathbf{u}^h(\mathbf{x}, t_k)
\end{equation}
is used as the base flow for the ansatz (\ref{eq:ansatz}). Finally, the snapshot set
\begin{equation}
	\mathcal{U}'' = \{ \mathbf{u}^h(\mathbf{x}, t_k) - \overline{\mathbf{u}}(\mathbf{x})\}_{k=1}^M
\end{equation}
is used for the POD algorithm. As it is common practice, the ``snapshot'' method of \citet{Sirovich1987} is used.

The selection of the number of basis functions used for the projection, and hence the dimension of the state vector $\mathbf{a}$, is driven by the trade-off between accuracy and cost of computations. The key aspect in this selection is that the computational cost associated with the solution of SOS problems (\ref{eq:mini-no-control-sos}, \ref{eq:mini-control-sos}) increases quite dramatically with the state dimension $N$ and the degrees of the function $V$ and the controller $g$, $d_V$ and $d_g$. For example, the SOS constraint in (\ref{eq:mini-no-control-sos}) is a polynomial in $N$ variables of degree $d_V+1 = 2d$, assuming that $\Phi$ has lower degree than this, and because $\mathbf{f}$ is at most quadratic in $\mathbf{a}$ for models of incompressible fluid flows. For such a polynomial, the vector of monomials equivalent to $\mathbf{v}$ in (\ref{eq:sos-quadratic-form}) consists of $D=(N+d)!/(N!d!)$ individual terms, whereas the cost for solving the SDP problem in each iteration increases in practice as $\mathcal{O}(D^3)$ (see \citet{Goulart2012692} and references therein for more details).

\begin{figure}
	\centering
	\includegraphics[width=0.85\textwidth]{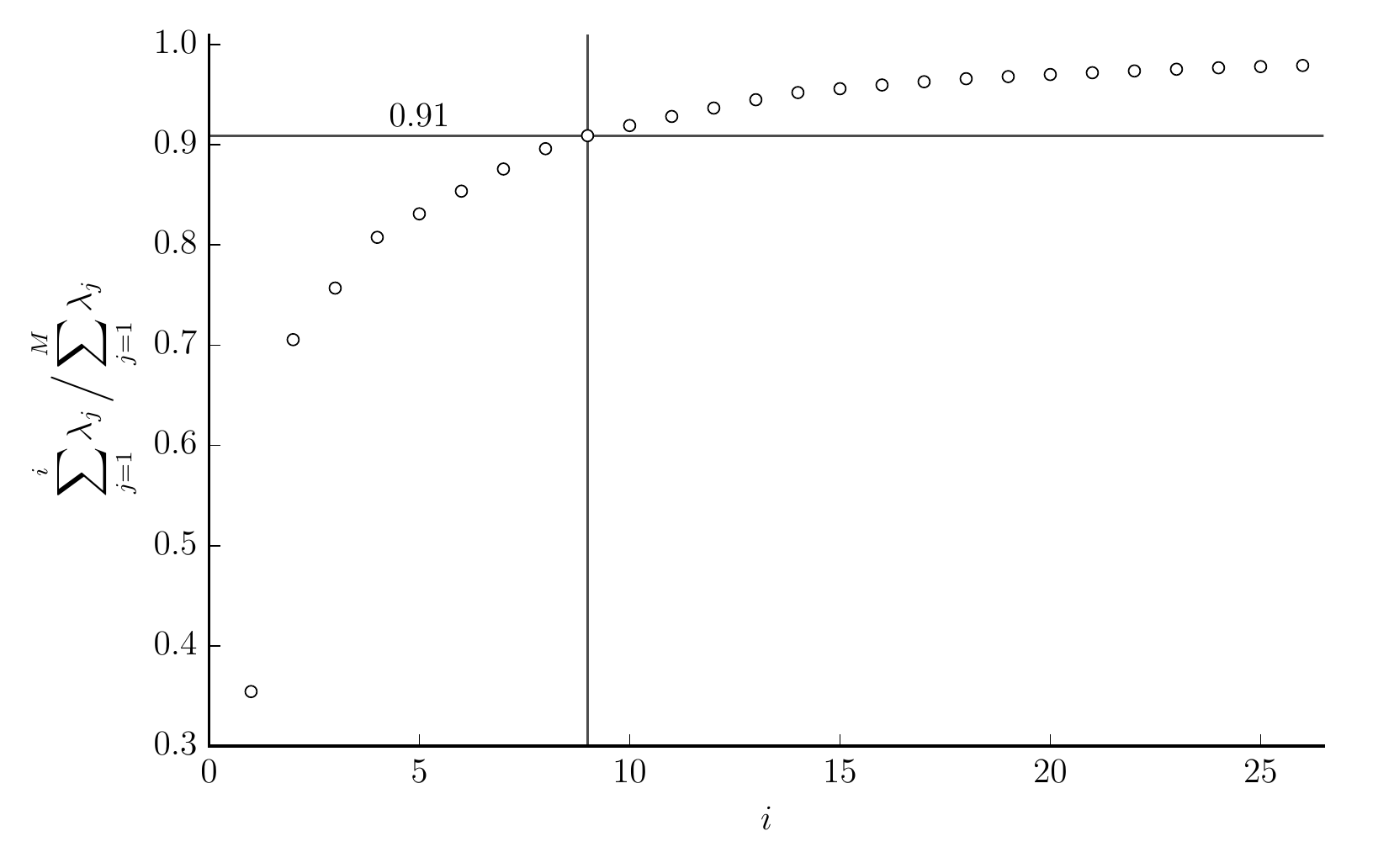}
	\caption{Normalised, cumulative energy associated with the POD modes obtained from snapshots sampled from DNS with random actuation. Nine POD modes are selected, capturing 91\% of the total fluctuation kinetic energy associated with the snapshots.}
	\label{fig:POD-spectrum}
\end{figure}
As a compromise between computational cost and model performance, we selected the first nine POD modes for the Galerkin projection, capturing about 91\% of the total fluctuation kinetic energy in the snapshots, as illustrated in figure \ref{fig:POD-spectrum}, which shows the normalised cumulative energy associated with the POD modes. In addition, this model is augmented with a tenth, shift mode \citep{noack-hierarchy} a particular mode spanning the direction from the mean flow $\overline{\mathbf{u}}(\mathbf{x})$ to the unstable, steady and symmetric solution $\mathbf{u}_0(\mathbf{x})$ of the equations (\ref{eq:ns}). The symmetric flow is obtained numerically as the steady-state solution using the half-upper grid of the original problem, with free-slip boundary condition on the symmetry plane, sufficient to suppress the symmetry-breaking unstable  normal mode that grows and saturates into the periodic von K\'arm\'an street \citep{bergmann2005optimal,Protas2002wt}. The shift mode is then constructed as
\begin{equation}
	\mathbf{u}_\Delta(\mathbf{x}) = \frac{\overline{\mathbf{u}} - \mathbf{u}_0}{\|\overline{\mathbf{u}}-\mathbf{u}_0\|}
\end{equation}
and it is then made orthogonal to the remaining nine POD modes using a Gram-Schmidt procedure.

It is well recognised \citep{tadmor2010meanfield} that the inclusion of shift modes in Galerkin models of natural and actuated wake flows past a circular cylinder improves transient dynamics over larger changes in base flow. However, a more important result is that inclusion of such a mode results in a dynamical system for which a finite upper bound on the long-time average of energy $\Phi(\mathbf{a}) = \mathbf{a}^T\mathbf{a}/2$ can be found using the procedure presented in section \ref{sec:bound-estimation}, similarly to what was observed in \citet{schlegel2015long}. On the other hand, the nine-mode POD-based system does not appear to have such a property, as we have been unable to find an upper bound for the same quantity. Even though the inability to find an upper bound using SOS does not prove that a finite upper bound does not exists, as discussed in section \ref{sec:bound-estimation}, it prevents the application of the control methodology proposed in this paper, entirely based on bounds estimation and optimisation.

Standard Galerkin projection is then performed by inserting the expansion (\ref{eq:ansatz}) in (\ref{eq:ns}), and setting the inner product with each of the modes to zero in turn. Neglecting the small contribution arising from the projection onto the pressure gradient field, as commonly done for this fluid flow (e.g. \citet{bergmann2005optimal,noack-hierarchy}), results in the nonlinear system of first-order coupled ordinary differential equations, the reduced-order model (ROM):
\begin{equation}\label{eq:ode-sys}
\displaystyle \frac{\mathrm{d}a_i}{\mathrm{d}t} = c_i + {\sum_{j=1}^N L_{ij} a_j}  + {\sum_{j=1}^N \sum_{k=1}^N Q_{ijk}a_j a_k} + {m_i   \frac{\mathrm{d}\gamma}{\mathrm{d}t}  + e_i \gamma + b_i \gamma^2 + \sum_{j=1}^N F_{ij} a_j \gamma} \quad i=1, \ldots N
\end{equation}
The definitions of the coefficients $c_i, L_{ij}, N_{ijk}, m_i, e_i, b_i, F_{ij}$ arising from the projection are standard and are reported in Appendix \ref{sec:app-ode-sys}. Numerical time integration of the ROM is performed using a standard fourth-order Runge-Kutta scheme with time step equal to $10^{-3}$.

\subsection{Choice of the cost function}
It has been pointed out in the literature \citep{homescu} that the choice of the quantity to be minimised by control can sometimes determine the performance of the resulting controller. Hence, several options have been proposed. For instance, in the optimal control approach of \citet{protas2002optimal}, using full-order simulations of the Navier-Stokes equations and their adjoint, the chosen cost was the sum of the work needed to resist the drag force and the work needed to control the flow. These two quantities could be computed exactly for that case, but for reduced-order Galerkin-type models, such level of detail is typically not available, or require extension of the POD basis to pressure,\citep{Bergmann2009}. In some works \citep{graham1999optimal, bergmann2005optimal}, the unsteadiness in the wake is typically used as a proxy for drag, and an additional penalisation on the control magnitude is added for regularisation purposes. In the present work, we adopted this formulation where the cost to be reduced is the domain integral of the kinetic energy of the velocity fluctuations resolved by the ansatz (\ref{eq:ansatz}), plus a penalisation on the control, i.e. the quantity
\begin{equation}\label{eq:cost}
	\Phi(\mathbf{a}(t)) = \frac{1}{2}\mathbf{a}(t)^T \mathbf{a}(t) + R \gamma^2( \mathbf{a}(t))
\end{equation}
where the orthonormality of the basis functions has been used. The penalisation factor $R$ does not have an immediate physical meaning, but it is used as a design parameter as a means to artificially limit the amplitude of the control. This is necessary because increasingly large control inputs will drive the ROM increasingly far from the region of the phase space where accurate and realistic dynamical behaviour can be expected. As a result, performance in DNS can be affected, as it will be shown later.


\subsection{Model calibration}
The ten-mode reduced-order model obtained directly from Galerkin projection is able to represent the dynamics of the full-order system only over a short time scale, i.e. about one shedding cycle, and the long-term behaviour is not correctly represented.
\begin{figure}
	\centering
	\includegraphics[width=0.92\textwidth]{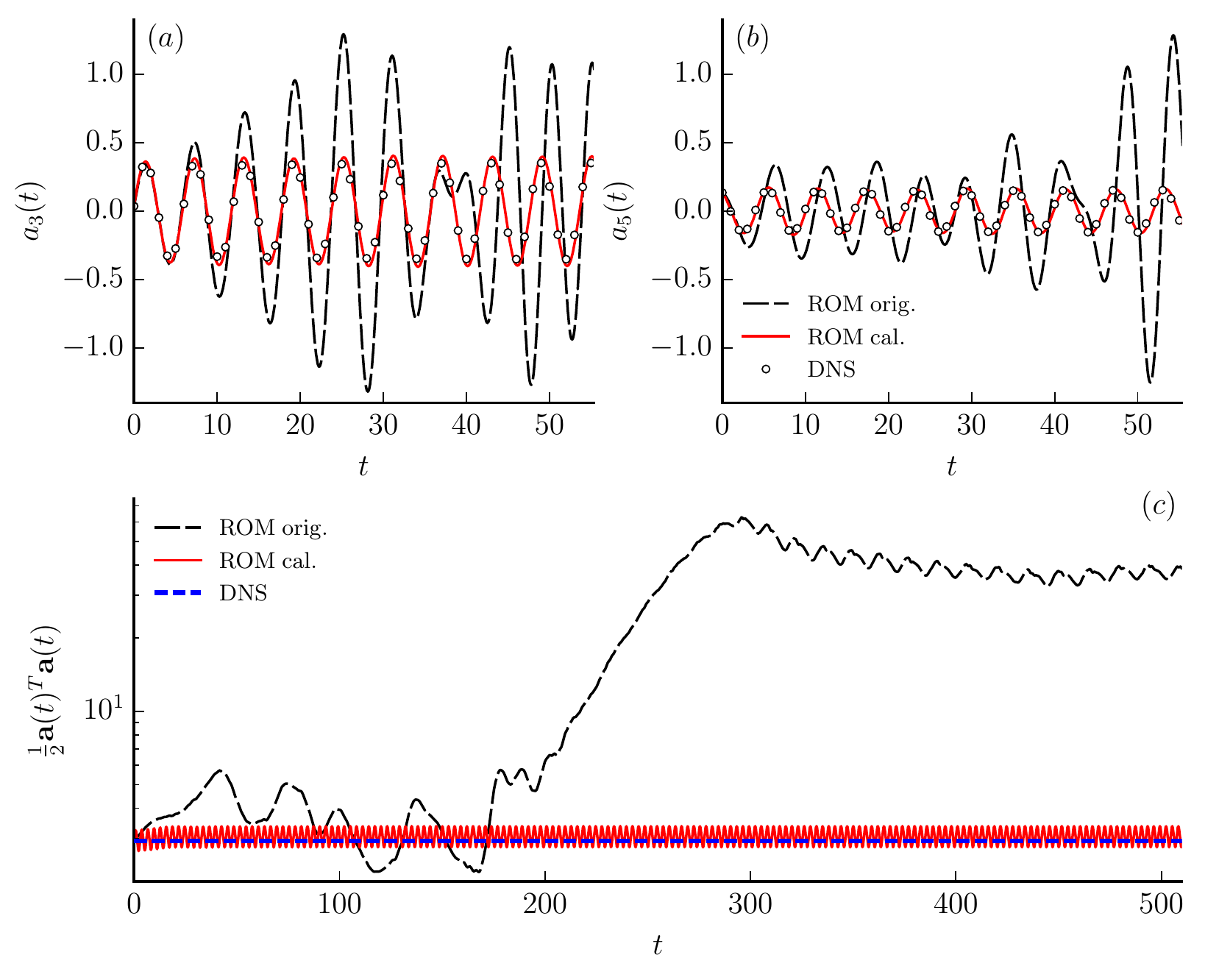}
	\caption{(Colour online). Panels $(a)$, $(b)$: time histories of states $a_3$ and $a_5$ from numerical integration of the ROM obtained directly from Galerkin projection, black dashed line, and of the calibrated ROM, red solid line, compared with the time history of the projections of the corresponding POD modes onto the DNS solution. Panel $(c)$: time histories of system energy for the original and calibrated ROMs, and from projections on the DNS of the uncontrolled flow.}
	\label{fig:calibration}
\end{figure}
This phenomenon is illustrated in figure \ref{fig:calibration}. Panels $(a)$ and $(b)$ show time histories of the projections of the third and fifth POD modes, respectively, onto the direct numerical simulation of the uncontrolled flow, i.e. the quantities
\textcolor{black}{\begin{equation}
\tilde{a}_i(t) = \langle \mathbf{u}_i(\mathbf{x}), \mathbf{u}(\mathbf{x}, t) - \overline{\mathbf{u}}(\mathbf{x})\rangle, \quad i =3, 5
\end{equation}}
These are compared with the time histories of the same quantities obtained from numerical integration of the original ROM, with initial condition $\mathbf{a}(0) = \tilde{\mathbf{a}}(0)$, black dashed line. It is clear that the predictions of the model quickly diverge and become essentially useless. The energy of the system $\mathbf{a}^T\mathbf{a}/2$, panel $(c)$, grows significantly and its long-time average is twelve times larger than the mean resolved energy obtained from projections of the modes onto the DNS solution. \textcolor{black}{The attractor of the ROM is thus significantly different from} the projection of the stable limit cycle associated with vortex shedding onto the ten-dimensional phase space. This is a recurrent problem in reducing the order of nonlinear dynamical systems \citep{Cordier2013-pc} because the high-sensitivity to perturbations, such as the truncation of low-energy modes, can have a profound effect after a sufficiently long time \citep{Marion1989}. A crucial point is that time-averages, and bound estimation/optimisation, depend strongly on the geometry of the phase space. It is then very important for an effective application of the proposed control design methodology to have a ROM matching as precisely as possible the long-term behaviour of the original full-order system.

Hence, we apply an eddy-viscosity model calibration scheme, which has become standard practice to correct the effects of unresolved, truncated modes \citep{couplet, sisirup, Noack_undated-uc, Bergmann2009, Protas2015-an}.  Following previous work \citep{cordier-calibration}, we add to (\ref{eq:sys}) a linear calibration term $L^c_{ij} a_j$, where the matrix $L^c_{ij}$ has non-zero, initially undetermined, entries only on the main diagonal and the first upper/lower diagonals. Adding the contribution from the two off diagonals, as opposed to previous work where only the diagonal elements are identified \citep{galletti2004low} was necessary to achieve satisfactory tracking of the reference limit cycle. Optimal entries are obtained from the solution of the optimisation problem
\begin{equation}\label{eq:optimisation-Lc}
\displaystyle \underset{L^c_{ij}}{\text{min}} \displaystyle \int_{t_0}^{t_1} \| \mathbf{a}(t; L^c_{ij}) - \tilde{\mathbf{a}}(t) \|^2 \mathrm{d}t
\end{equation}
subject to the state equation (\ref{eq:ode-sys}) with $\gamma(t)=0$, with initial condition $\mathbf{a}(t_0) = \tilde{\mathbf{a}}(t_0)$. In (\ref{eq:optimisation-Lc}), the time integral of the norm of the error between the calibrated ROM trajectory $\mathbf{a}(t; L^c_{ij})$ and the projection of the trajectory of the full-order system onto the selected subspace $\tilde{\mathbf{a}}(t)$ is minimised. This trajectory is obtained from numerical simulation of the uncontrolled flow, after transients have died out, in order to force the calibrated ROM to describe correctly the stable limit cycle associated with vortex shedding. A sequence of optimisation problems with increasing $t_1-t_0$ is formulated to improve convergence of this non-convex problem, with the final interval amounting to about 20 shedding cycles. This procedure is not guaranteed to result in successful identification in the general case, but was successful in this case.

Once calibrated, the ROM shows a more realistic behaviour, as the time-averaged energy on the attractor is only 4\% greater than that associated with the limit cycle of the full-order system, as illustrated in figure \ref{fig:calibration} (solid red lines). However, poor controllability was observed, as opposed to larger models that do not presented this behaviour, suggesting that the rotary actuation of the cylinder affects via viscosity the large scale motions, i.e. the resolved modes, through linear/nonlinear interaction with the truncated modes. To mitigate this poor controllability, we added two additional calibration terms $e^c_i a_i$ and $m^c_i \mathrm{d}\gamma/\mathrm{d}t$ in (\ref{eq:ode-sys}). Optimal values are obtained from an optimisation problem similar to (\ref{eq:optimisation-Lc}), where the numerical simulation used to determine the POD modes is used as reference. Although the trajectory of the ROM remains bounded when integrated using the same actuation signal of DNS, it quickly diverges from the reference trajectory and rapidly becomes uncorrelated. Hence, we adopted a more robust multiple-shooting identification scheme \citep{peifer, Protas2015-an}. The idea is to consider a set of $K$ blocks, each of length $T$ equal to the shedding period and minimise the sum of all deviations, i.e. solving
\begin{equation}\label{eq:optimisation-ejmj}
\displaystyle \underset{e_i^c, m_i^c}{\text{min}} \displaystyle \sum_{k=1}^K \int_{t_k}^{t_k + T} \| \mathbf{a}(t; e_i^c, m_i^c) - \tilde{\mathbf{a}}(t) \|^2 \mathrm{d}t,
\end{equation}
subject to the state equation (\ref{eq:ode-sys}), with identical notation as in (\ref{eq:optimisation-Lc}), and where the initial condition $\mathbf{a}(t_k) = \tilde{\mathbf{a}}(t_k)$ is used for numerical integration of the calibrated ROM over the $k$-th block.

\section{Results}
The SOS-based methodology discussed above has been used to derive a set of linear state-feedback controllers, i.e. the degree $d_g$ has been set to one, with $d_V = 4$ in all cases, for various penalisation factors. Additional tests have been performed with $d_V=6$ with no difference, except for additional computational costs, as all bounds are tight to the actual average from simulation with $d_V=4$. Linear controllers have the form \textcolor{black}{\mbox{$\gamma(t) = \sum_{i=1}^N k_i a_i(t)$}}, where all the gains have to be identified. \textcolor{black}{The constant term is manually set to zero, i.e. the gains $k_i$ are the only decision variables in the SOS calculations. This is done explicitly to avoid naturally occurring spurious control solutions with large non zero mean rotation, a result likely exploiting unrealistic dynamics described by the ROM far away from the operating regime.} It is possible to get rid of the term $m_i\mathrm{d}\gamma/\mathrm{d}t$ in (\ref{eq:ode-sys}), by noting that \textcolor{black}{$\mathrm{d}\gamma/\mathrm{d}t =\sum_{i=1}^N k_i\mathrm{d}a_i/\mathrm{d}t$ } and using the state equation to get
\begin{equation}\label{eq:trick}
	\displaystyle \mathrm{d}\gamma/\mathrm{d}t = \frac{1}{{1-\displaystyle \sum_{l=1}^N k_lm_l}}\sum_{i=1}^N k_i\Big(c_i + L_{ij} a_j  + N_{ijk}a_j a_k + e_i \gamma + b_i \gamma^2 + F_{ij} a_j \gamma \Big)
\end{equation}
which is substituted back in the state equation (\ref{eq:ode-sys}). This method cannot be used for nonlinear controllers, as the denominator of the fraction in (\ref{eq:trick}) would contain an expression in $\mathbf{a}$, making the resulting system non-polynomial in the state variables. A different approach is required as described in appendix \ref{sec:nonlinear-controllers}. As a matter of fact, the degree of the polynomial-type feedback controller can be regarded a design parameter, and it is just for the sake of simplicity that we consider linear controllers only in this paper, leaving the derivation and testing of nonlinear controllers as future work. It is worth pointing out that even though the feedback is a linear function of the state, the control design process is aware and exploits the fully nonlinear dynamics of the ROM. No linearisation is performed, unlike in \citet{AleksicRoesner2013ey}, who studied a similar feedback control configuration.

Feedback control results are reported for the ROM first, and then for the implementation in direct numerical simulation.

\subsection{Bound estimation and optimisation}\label{sec:linear-controllers}
An estimate of the long-time averaged cost (\ref{eq:cost}) was obtained by long numerical integrations of the ROM without control, starting from several random initial conditions. All trajectories converged to the same stable limit cycle and the associated long-time averaged cost was $\overline{\Phi}^0 = 3.07$. Trajectories never exhibited blow up, nor converged to a different attractor, although these numerical experiments cannot, of course, be considered as a proof that another stable attractor does not exist in the phase space of the ROM.

The estimation of the upper bound via SOS is performed by trial-and-error. For a given $C$ we try to find $V$ satisfying the constraint in (\ref{eq:mini-no-control-sos}). If this is successful in the sense that the resultant SOS decomposition satisfies the feasibility-checking condition (\ref{post-check}) \citep{Lofberg2009-mu} we decrease $C$ by $\delta C$, which is 0.01 here, and repeat the trial.
In checking the feasibility of the problem, it is important to consider that SOS problems such as (\ref{eq:mini-no-control-sos}) lead quickly to large semidefinite programmes, typically becoming strongly ill-conditioned as the size increases, although the numerical algorithms are based on convex programming. As discussed in detail in \citet{Lofberg2009-mu}, the equality constraints associated with (\ref{eq:sos-quadratic-form}), are only satisfied in the limit of the solution process, as a result of finite-precision arithmetic and various termination criteria.



\begin{figure}
	\centering
	\includegraphics[width=0.95\textwidth]{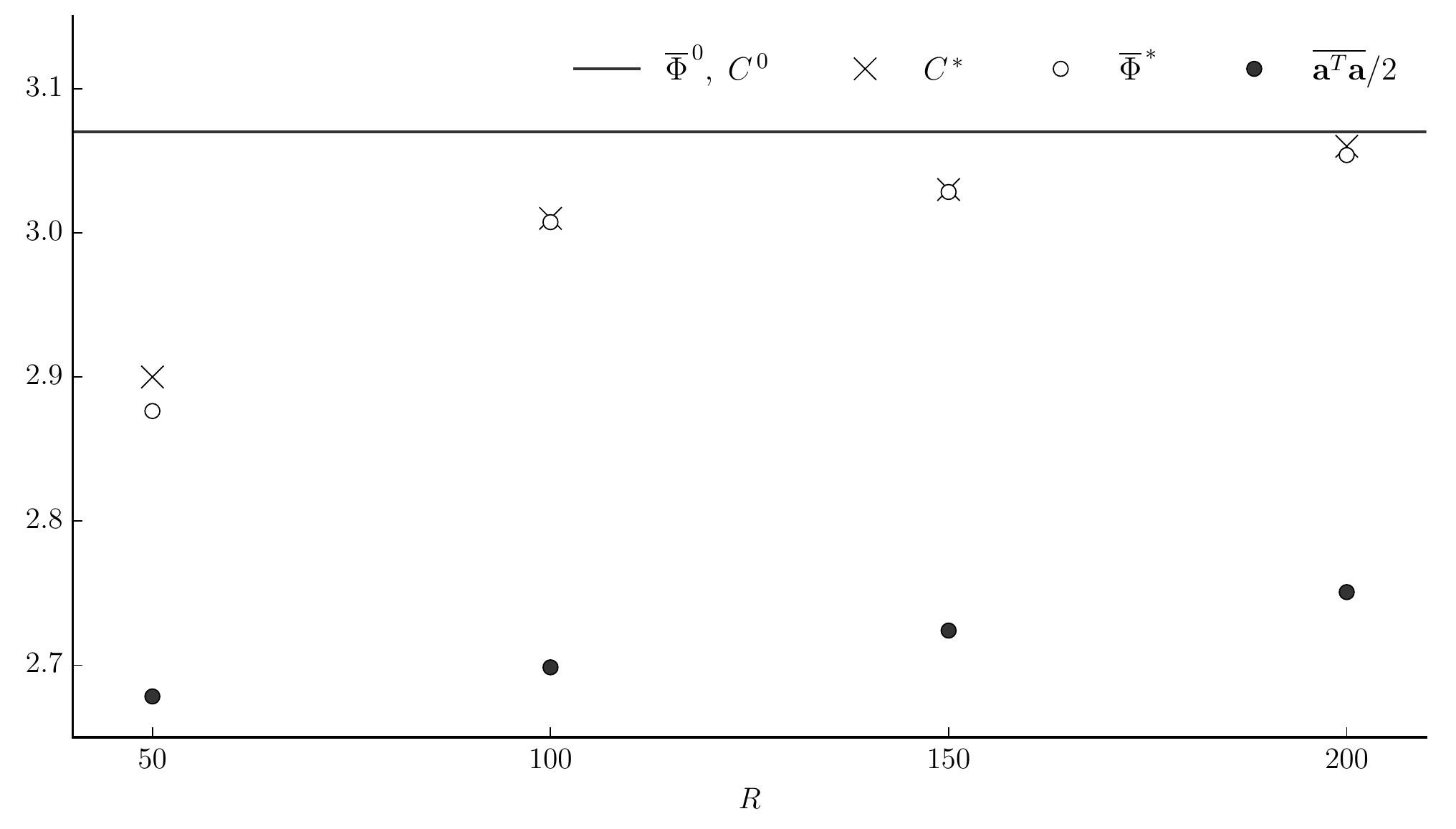}
	\caption{(Colour online). Performance of linear feedback controllers for various penalisation factors $R$ in closed-loop simulation of the ROM. Crosses ($\times$): upper bound of the long-time averaged cost; open circles ($\ocircle$): converged value of the long-time averaged cost; closed symbols (\newmoon): long-time average of the resolved fluctuation kinetic energy. The horizontal line denotes the time average/upper bound for the uncontrolled system.}
	\label{fig:performance-rom}
\end{figure}

Several linear controllers have been designed for increasing values of the penalisation factor $R$. \textcolor{black}{Large values of $R$ have been used as these lead to better performance in DNS.} In figure \ref{fig:performance-rom} the performance of these controllers in closed-loop simulation of the ROM is summarised. Long-time averages of the cost are computed from numerical simulations started from an initial condition on the ROM's limit cycle and by discarding initial transients as control is activated. The figure reports the upper bound $C^*$ (crosses) the actual time average $\overline{\Phi}^*$ from simulation (open symbols) and the long-time average of the resolved fluctuation kinetic energy $\mathbf{a}^T\mathbf{a}/2$ (closed symbols) with the difference between the two latter quantities being the average cost of control. The horizontal line is the value for the uncontrolled system. Numerical values of the points displayed in figure \ref{fig:performance-rom} are also reported in table \ref{tab:table-results}, together with other quantities of interest.
The SOS-based control design successfully reduces the upper bound of the system. The reduction is larger for small $R$, as larger control magnitude are allowed, as it can be deduced by the last column of table \ref{tab:table-results}, which shows the root-mean-square value of the control input. The maximum reduction of the bound is relatively small, i.e. about 6\% for $R=50$; larger reduction can be found for smaller penalisation factors, although these controllers performed poorly in DNS.  A significant part of the total time-averaged cost comes, artificially, from the control. In fact, the time-averaged resolved fluctuation kinetic energy decreases by as much as 13\% for $R=50$ and by 11\% for $R=200$.

Repeated integration of the controlled ROM, from several random initial conditions, shows that the time average $\overline{\Phi}^*$ is always below the upper bound, and no instability of the closed-loop system has been observed. The upper bound is tight to the actual average, within the uncertainty of the numerics involved in solution of the SOS problems, as for the estimation of the bound for the uncontrolled case. The upper bound and the average from simulation appear to converge asymptotically to the bound of the uncontrolled system as $R$ increases.

\begin{figure}
	\centering
	\includegraphics[width=1\textwidth]{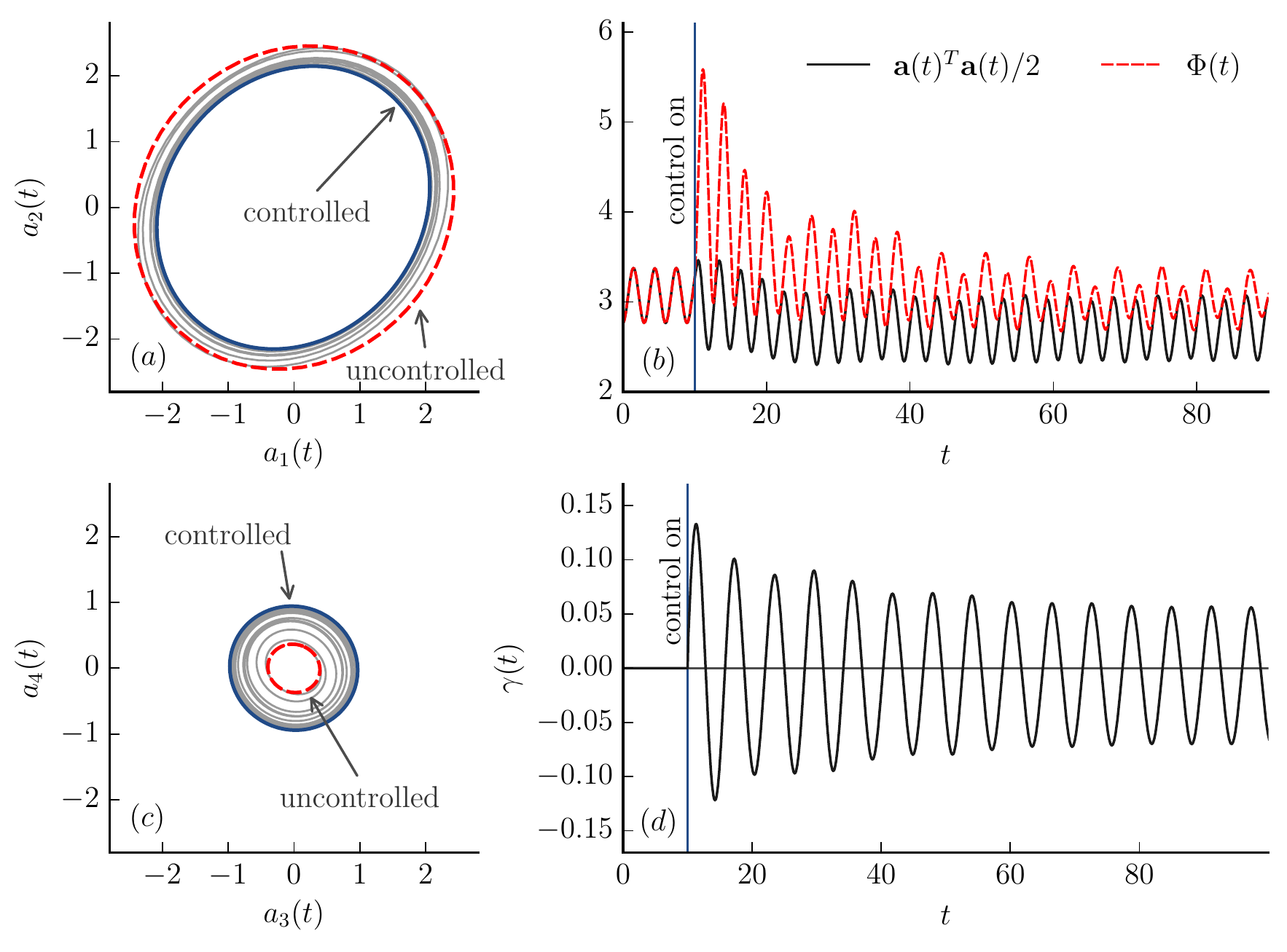}
	\caption{(Colour online). Transient dynamics of the controlled ROM for $R=150$. \textcolor{black}{Panels $(a)$ and $(c)$} show the trajectory of the ROM projected onto two relevant subspaces. The long-term behaviour of the system is indicated by the uncontrolled and controlled limit cycles. Panel $(b)$ shows time histories of the total cost and of the resolved fluctuation kinetic energy. Panel $(d)$ shows the time history of the control input.}
	\label{fig:transient-rom}
\end{figure}

Figure \ref{fig:transient-rom} shows the effects of control on the dynamics of the ROM, for the linear controller with $R=150$, reported as an example as the other controllers have a qualitatively similar impact on the dynamics. Panels $(a)$ and $(c)$ show the trajectory of the ROM projected in the $(a_1, a_2)$ and $(a_3, a_4)$ planes, respectively. The red/blue ``controlled''/``uncontrolled'' orbits denote the limit cycle before/after the activation of control. The transient between the two is indicated in light grey. The actuated dynamics converge to a controlled limit cycle, over which the mean resolved kinetic energy is reduced. Under the action of control, the energy of the first two modes decreases, whereas that of modes $a_3$, $a_4$ increases slightly. Physically, such a shift is interpreted as a restructuring of the wake as both pairs of modes correspond spatially to velocity fluctuations oscillating at the shedding frequency.
Panel $(b)$ shows time histories of the resolved energy (solid black line) and of the total cost (dashed red line). Feedback control is started at $t=10$. As soon as control is activated, the total cost $\Phi$ jumps up to about 5.5, because the control input, shown in panel $(c)$ quickly jumps to about 0.15. As anticipated, the penalisation in the control in the cost function does not limit the instantaneous value of the control, as a hard saturation would, but only its long-time averaged contribution.

The resolved kinetic energy decreases substantially in a short transient that takes about 10 time units, i.e. just less than two shedding cycles. On the other hand, the total cost takes a longer time to settle to the steady state, approximately 70 time units after activation of control, because the peak-to-peak variation of the control input $\gamma(t)$ decreases slowly during the transient.

\begin{table}
  \begin{center}
\def~{\hphantom{0}}
  \begin{tabular}{rlllll}
      $R$  & $C^*$ & $\overline{\Phi}^*$ & $\overline{\mathbf{a}^T\mathbf{a}}/2$  & $R\overline{\gamma^2}$ & $\sqrt{\overline{\gamma^2}}$ \\[3pt]
       50           & 2.88  & 2.876           & 2.678                                           & 0.198 &  0.063\\
       100          & 3.01  & 3.008           & 2.699                                           & 0.309 &  0.056\\
       150          & 3.03  & 3.028           & 2.724                                           & 0.304 &  0.045\\
       200          & 3.06  & 3.054           & 2.751                                           & 0.303 &  0.039\\
       uncontrolled & 3.07  & 3.070           & 3.070                                           &  &  \\
  \end{tabular}
  \caption{Linear feedback control results for the ROM for different penalisation factors $R$.}
  \label{tab:table-results}
  \end{center}
\end{table}

\subsection{Feedback control in DNS}
The four linear controllers derived are implemented in direct numerical simulation. Because the governing equations are marched forward in time, at the beginning of the $k$-th time step, at time $t_k$, the current state vector is obtained from the projections of the POD modes on the current fluctuating velocity field as
\begin{equation}\label{eq:projections-dns}
a_i(t_k) = \langle \mathbf{u}_i(\mathbf{x}), \mathbf{u}(\mathbf{x}, t_k) - \overline{\mathbf{u}}(\mathbf{x}) - \gamma(t_k) \mathbf{u}_c(\mathbf{x})\rangle, \quad i = 1, \ldots, N.
\end{equation}
The control action $\gamma(t_k)$, the tangential velocity on the cylinder surface, is calculated from the control law (\ref{eq:sos-quadratic-form}) and is then set as constant boundary condition for the time step $t_{k+1}-t_k$.

\begin{figure}
	\centering
	\includegraphics[width=1\textwidth]{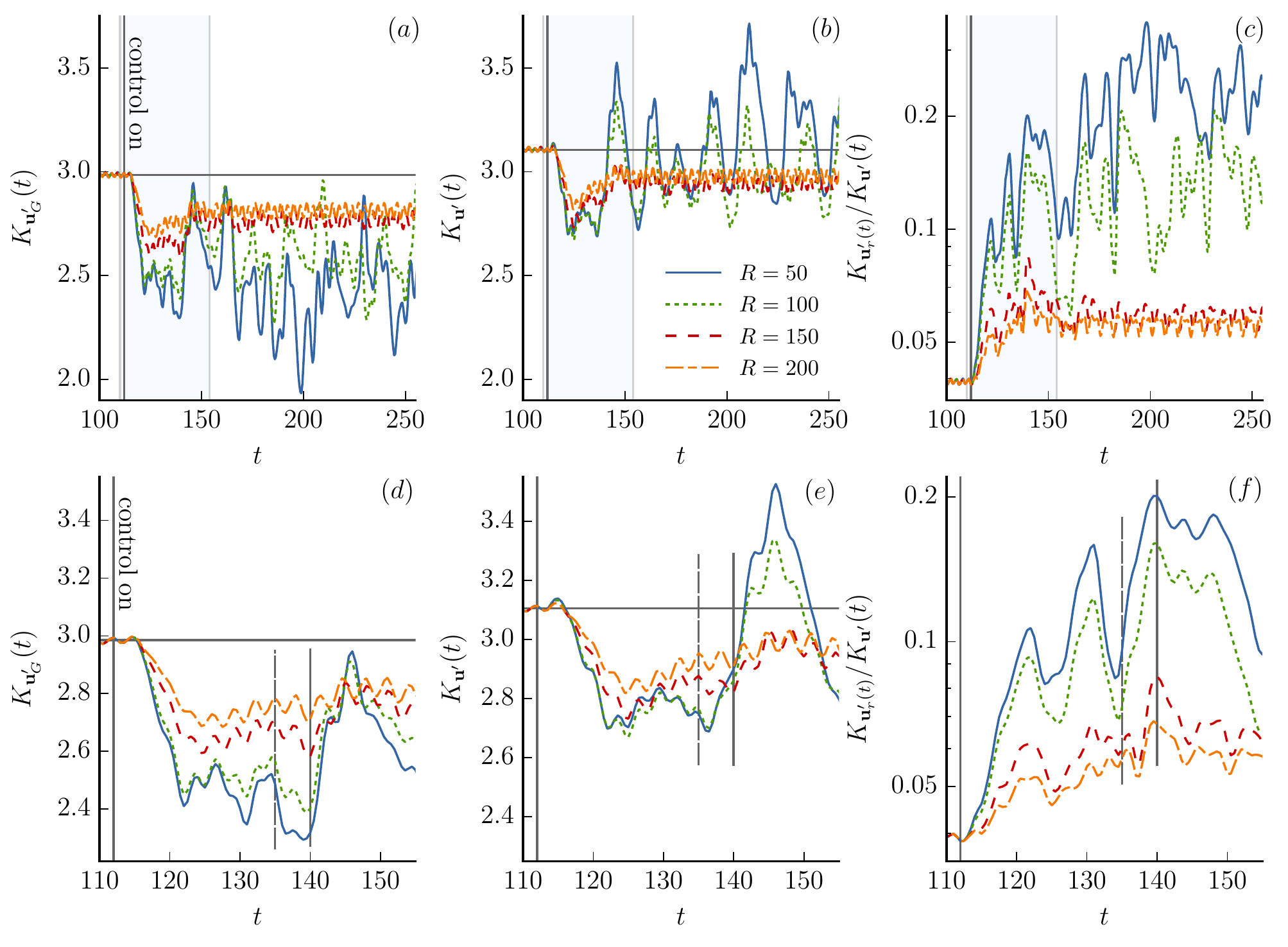}
	\caption{(Colour online). Performance of linear feedback controllers in DNS. Panel $(a)$: time history of fluctuations kinetic energy resolved by the ansatz (\ref{eq:ansatz}); panel $(b)$: time history of the total fluctuation kinetic energy; panel $(c)$: time history of the unresolved residual energy, normalised with the total fluctuation kinetic energy. The bottom panels show the same quantities in the interval $t \in [110, \,155]$.}
	\label{fig:resolved-and-total-energy}
\end{figure}

Figure \ref{fig:resolved-and-total-energy} shows time histories of three key quantities obtained from direct numerical simulation of the closed-loop system.  These are the fluctuation kinetic energy resolved by the Galerkin expansion (\ref{eq:ansatz})
\begin{equation}
	K_{\mathbf{u}'_G}(t) = \frac{1}{2}\| \mathbf{u}'_G(\mathbf{x}, t)\|^2  =  \frac{1}{2}\| \sum_{i=1}^N \tilde{a}_i(t)\mathbf{u}_i(\mathbf{x})\|^2 = \frac{1}{2}\tilde{\mathbf{a}}(t)^T \tilde{\mathbf{a}}(t),
\end{equation}
obtained from the projections (\ref{eq:projections-dns}), in panel $(a)$; the total fluctuation kinetic energy
\begin{equation}
	K_{\mathbf{u}'}(t) = \frac{1}{2}\|\mathbf{u}(\mathbf{x}, t) - \overline{\mathbf{u}}(\mathbf{x}) - \gamma(t)\mathbf{u}_c(\mathbf{x})\|^2,
\end{equation}
in panel $(b)$; and the kinetic energy of the residual fluctuation $\mathbf{u}_r(\mathbf{x}, t) = \mathbf{u}'(\mathbf{x}, t) - \mathbf{u}'_G(\mathbf{x}, t)$
\begin{equation}
	K_{\mathbf{u}_r'}(t) = K_{\mathbf{u}'}(t) - K_{\mathbf{u}'_G}(t),
\end{equation}
normalised with the total fluctuation kinetic energy $K_{\mathbf{u}'}$, in panel $(c)$. Note that the cost of the control $R\gamma(t)^2$ is not added to panel $(a)$.
The bottom panels show the same quantities in the interval $t \in [110, \;155]$, in the initial transient after activation of feedback control at $t=112$.

\textcolor{black}{As soon as control is activated,} the resolved and total kinetic energy decrease substantially, in a transient lasting for about 10-12 times units, similarly to what exhibited by the ROM in figure \ref{fig:transient-rom}. The initial time rate of change of the energy and the maximum reduction are larger for smaller penalisation factors, as the control is more aggressive.  Subsequently, the cost remains approximately constant for a short period, in the interval $t \in [125, \;140]$. For $R=200$, $K_{\mathbf{u}'_G}$ has an average value in this window, roughly equal to that obtained in simulation of the ROM, in figure \ref{fig:performance-rom}. For the lower penalisation factors, the reduction of the resolved energy is larger than what obtained in simulation of the ROM, suggesting that the ROM significantly underestimates the effects of control on the dynamics. The reduction of the resolved and, most importantly, of the total fluctuation energy suggests that control design has successfully identified the control mechanism to attenuate vortex shedding.

In panel $(f)$, the fraction of unresolved kinetic energy grows significantly in this first interval, from a value of about 4\%, up to about 20\% for the smaller $R$. This shows that under the effects of control the full-order system explores regions of the high-dimensional phase space not included in the initial low-dimensional subspace chosen for the projection, especially for larger control inputs. \textcolor{black}{Taking into account the slow deformation of the wake structure unaccounted for in the original POD basis, using, e.g., deformable modes \citep{tadmor-rs} or updating the modes set (\citet{Bergmann2009} and references therein) might be beneficial to limit this behaviour and achieve improved performance.}

After these initial stages, the character of the solution depends strongly on the penalisation $R$. For $R=150, 200$, the long-term behaviour of the system is a controlled limit cycle with a reduced fluctuation kinetic energy, as predicted by the ROM in figure \ref{fig:transient-rom}. By contrast, for the two lower penalisation factors, the structure of the long-term behaviour is significantly different. The time history of the total fluctuation kinetic energy, panel $(b)$, undergoes a periodic low-frequency, large-amplitude modulation, with a period of about 18 time units, not clearly visible from the resolved energy in panel $(a)$.

\begin{figure}
	\centering
	\includegraphics[width=1\textwidth]{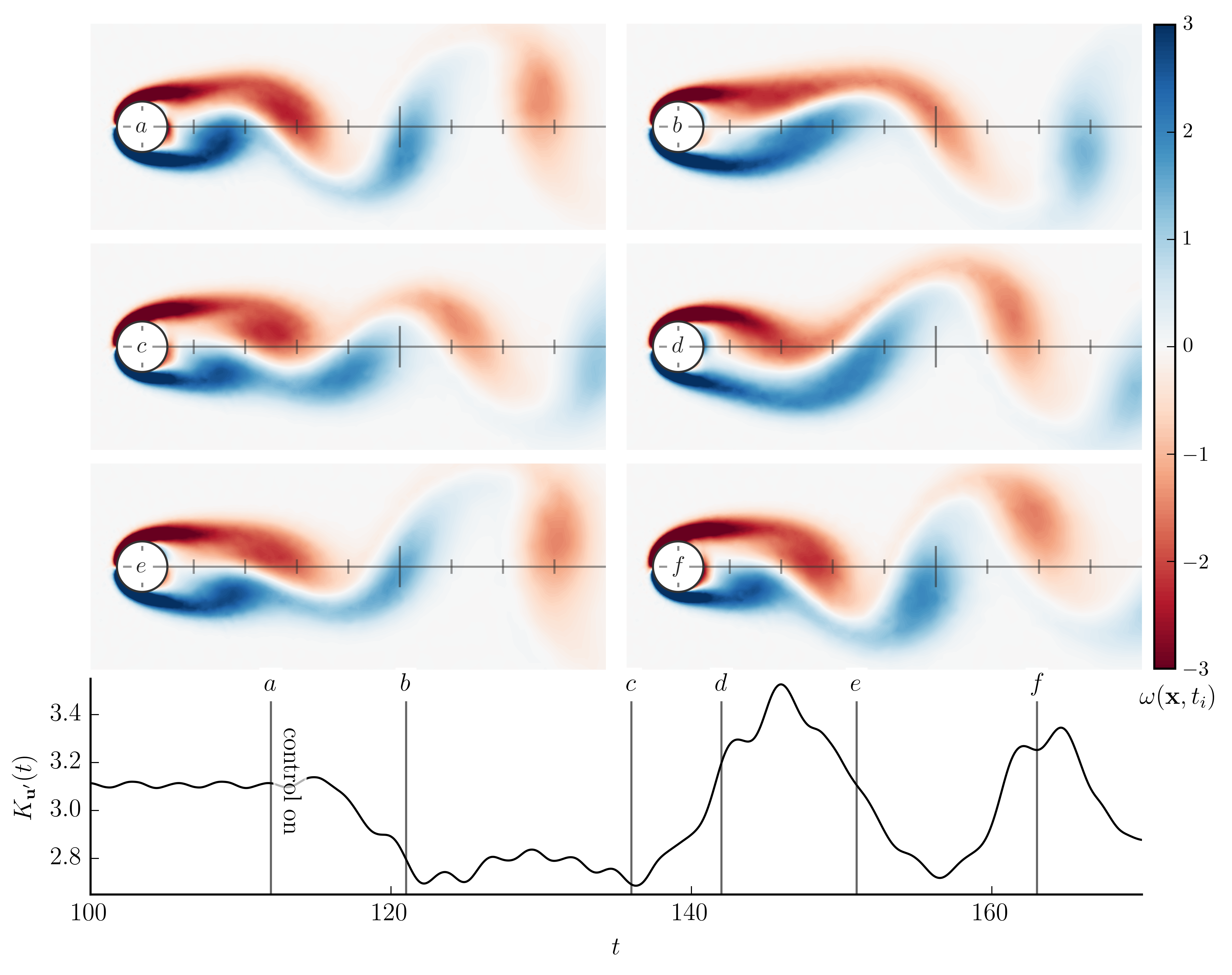}
	\caption{(Colour online). The top six panels show snapshots of vorticity from direct numerical simulation of the controlled flow with linear controller with $R=50$. The bottom panel shows the time history of the total fluctuation kinetic energy. Vertical lines denote the time instant at which snapshots are extracted, at $t=112, 121, 136, 142, 156$ and $162$.}
	\label{fig:snapshots-50}
\end{figure}

Insight into this phenomenon can be gained by analysing in more detail the behaviour of the system from about $t=135$ onwards, indicated with a dashed vertical segment in the bottom panels. The total kinetic energy, panel $(e)$, and the normalised unresolved energy, panel $(f)$, increase significantly for $R=50$ and 100. The resolved fluctuation energy is practically constant in the interval $t \in [135, \;140]$, panel $(d)$, and only when this growth saturates the quantity $K_{\mathbf{u}'_G}$ begins to increase. The physical mechanism responsible of this behaviour is illustrated in figure \ref{fig:snapshots-50}, for $R=50$. The figure shows six snapshots of the vorticity field, with the colour map clipped at $\pm3$ to visualise the structure of the actuated wake, although the maximum vorticity magnitude can be as high as 25 in the boundary layer on the cylinder. The bottom panel shows the time history of the total fluctuating kinetic energy, also reported in panel $(b)$ of figure \ref{fig:resolved-and-total-energy}. The vertical lines denote the times $t_i$ at which the snapshots are extracted.

In the initial transient after activation of control, between $t=112$ and $t=121$ (snapshots $a$ and $b$) the controlled rotation of the cylinder reorganises the generation and dynamics of vorticity in the near wake. The roll-up of the two shear layers is delayed and the fluctuation energy decreases steadily. Shortly after time $t_b$, the fluctuation energy stops decreasing and during the interval $[t_b, \; t_c]$ the wake locks onto an actuated limit cycle, with reduced $K_{\mathbf{u}'}$. The structure of the wake in this regime, panel $(c)$, is significantly different from the unactuated wake of panel $(a)$. It is narrower, especially in $4 \lesssim x \lesssim 9$, and the streamwise separation between the structures is shorter. Shortly after $t_c=136$, the fluctuation kinetic energy grows rapidly. This event is connected to the break down of the wake structure of panel $(c)$, arising as a large scale flapping of the entire near wake flow. This effect might be connected the growth of an instability of this wake structure, or it might be simply induced by the feedback. After $K_{\mathbf{u}'}$ peaks, the restructuring of the wake into the original uncontrolled state enables the control to reduce the fluctuation energy, panel $(e)$, although the same break-up observed in snapshot $(c)$ occurs shortly after $t_e$. This mechanism repeats indefinitely originating the low-frequency modulation visible in figure \ref{fig:resolved-and-total-energy}.

\begin{figure}
	\centering
	\includegraphics[width=1\textwidth]{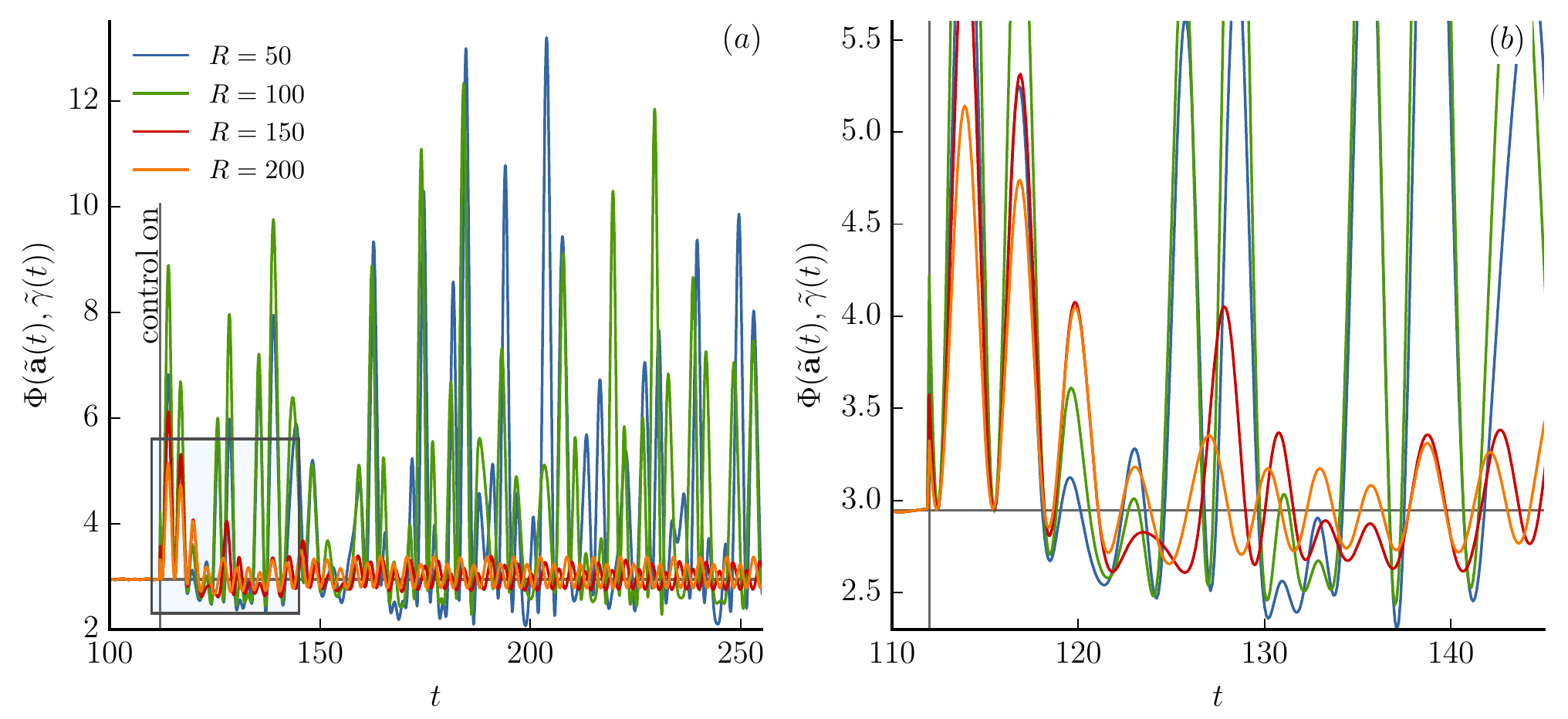}
	\caption{(Colour online). Time history of the total cost $\Phi$, as defined in equation (\ref{eq:cost}). Panel $(b)$ shows a detail of the same time trace, in a small time interval at the early stages of the simulation, indicated by the rectangle in panel $(a)$.}
	\label{fig:total-cost}
\end{figure}

Although the total fluctuation kinetic energy is successfully reduced in direct numerical simulation of the closed-loop system, the long-time average of the total cost $\Phi(\tilde{\mathbf{a}}(t), \tilde{\gamma}(t))$, as defined in (\ref{eq:cost}), is not decreased. This result is shown in panel $(a)$ of figure \ref{fig:total-cost}, whereas panel $(b)$ contains a zoom of the same quantity at the initial stages of the simulation, when control is activated.
The total cost initially spikes at quite large values, as the control input is quite intense, similarly to what is observed for the ROM in figure \ref{fig:transient-rom} for $R=150$. As control modifies the wake structure, the fluctuations of the cost decrease substantially, below the reference value of the uncontrolled system (horizontal line) approximately in the interval $t \in [120,\;125]$. However, the loss of control performance described above in figure \ref{fig:snapshots-50} results in a strong increase of the total instantaneous cost, especially for the two lower penalisation factors. As a result, the long-time averaged cost $\overline{\Phi}^*$, reported in table \ref{tab:table-results-dns}, is above the reference value of the uncontrolled system, $\overline{\Phi}^0=2.95$, also for the two larger penalisation factors.

The ROM results, table \ref{tab:table-results}, show that, for $R=200$, the percentage reduction of the cost is quite small, about 0.5\%, as a rather large contribution comes from the control cost. In DNS, the same controller results in an increase of the total cost of about 5\%. The discrepancy between these two values is certainly within the accuracy of the ROM in describing the effects of actuation on the full-order dynamics.

A physically meaningful quantity is the long-time average of the total power spent to sustain the motion of the cylinder \citep{Bergmann2006ka}. This quantity, expressed per unit of span, is the sum of the power $P_D$ spent to move the cylinder at speed $u_\infty$ against the drag force $D$ and the power required to control the flow, i.e. the power $P_M$ required to rotate the cylinder at angular speed $\dot{\theta}$ (positive when counter-clockwise) against the viscous torque $M$ exerted by the fluid on the cylinder (positive when it induces a clockwise rotation.)
\begin{table}
  \begin{center}
\def~{\hphantom{0}}
  \begin{tabular}{rccccccccc}
      $R$  & $\overline{\Phi}^*$ & $\displaystyle 1/2\overline{\mathbf{a}^T\mathbf{a}}$  & $R\overline{\gamma^2}$ & $\sqrt{\overline{\gamma^2}}$  & $\overline{C_D}$ & PSR \\[3pt]
                50  &  4.0  & 2.53 & 1.45  & 0.170 & 1.343 & 2.5 \\
                100 &  4.4  & 2.58 & 1.82  & 0.135 & 1.335 & 6 \\
                150 &  2.97 & 2.73 & 0.24 & 0.040 & 1.338 & 75 \\
                200 &  3.04 & 2.78 & 0.27 & 0.036 & 1.348 & 78 \\
       uncontrolled &  2.95 & 2.95 & 0.00 & 0.000 & 1.400 & 0\\
  \end{tabular}
  \caption{Linear feedback control results in DNS for different penalisation factors $R$. For the uncontrolled system \textcolor{black}{$\overline{\Phi}^* = \overline{\Phi}^0 = 1/2 \overline{\mathbf{a}^T\mathbf{a}}$}}
  \label{tab:table-results-dns}
  \end{center}
\end{table}

The drag and viscous torque are determined by the dimensional pressure and viscous stress surface distributions $p(\theta)$ and $\tau(\theta)$ as
\textcolor{black}{
\begin{equation}
	D = -\int_{0}^{2\pi} p(\theta)\cos(\theta) + \tau(\theta)\sin(\theta)\mathrm{d}\theta, \quad M = -\frac{\mathcal{D}}{2}\int_{0}^{2\pi} \tau(\theta) \mathrm{d}\theta,
\end{equation}}
where the viscous stress on the surface arises from the distribution of the tangential velocity $u_\theta$ as
\begin{equation}
\tau(\theta) = \mu \frac{\partial u_\theta}{\partial r}\Big|_{\mathcal{D}/2},
\end{equation}
where $\mu$ is the dynamic viscosity.

The first contribution to the total power spent is then simply
\begin{equation}
	P_D = D u_\infty = \frac{1}{2} \rho {u_\infty}^3 \mathcal{D} C_D,
\end{equation}
whereas the second reads as
\begin{equation}
	P_M = M \dot{\theta} = 2M \gamma u_\infty \mathcal{D} = \rho {u_\infty}^3 \mathcal{D} C_M\gamma, 
\end{equation}
in which $C_D$ and $C_M$ are the coefficients of drag and moment, and $\gamma$ is the normalised surface velocity as introduced above.  Note that $P_M$ is positive when the cylinder transfers energy to the fluid and negative otherwise. In nondimensional terms, the total power spent is then expressed by the total power coefficient
\begin{equation}\label{eq:budget}
	\displaystyle C_P = \frac{P_D + P_M}{1/2\rho u_\infty^3 \mathcal{D}} = C_D + 2 C_M \gamma.
\end{equation}

\begin{figure}
	\centering
	\includegraphics[width=1\textwidth]{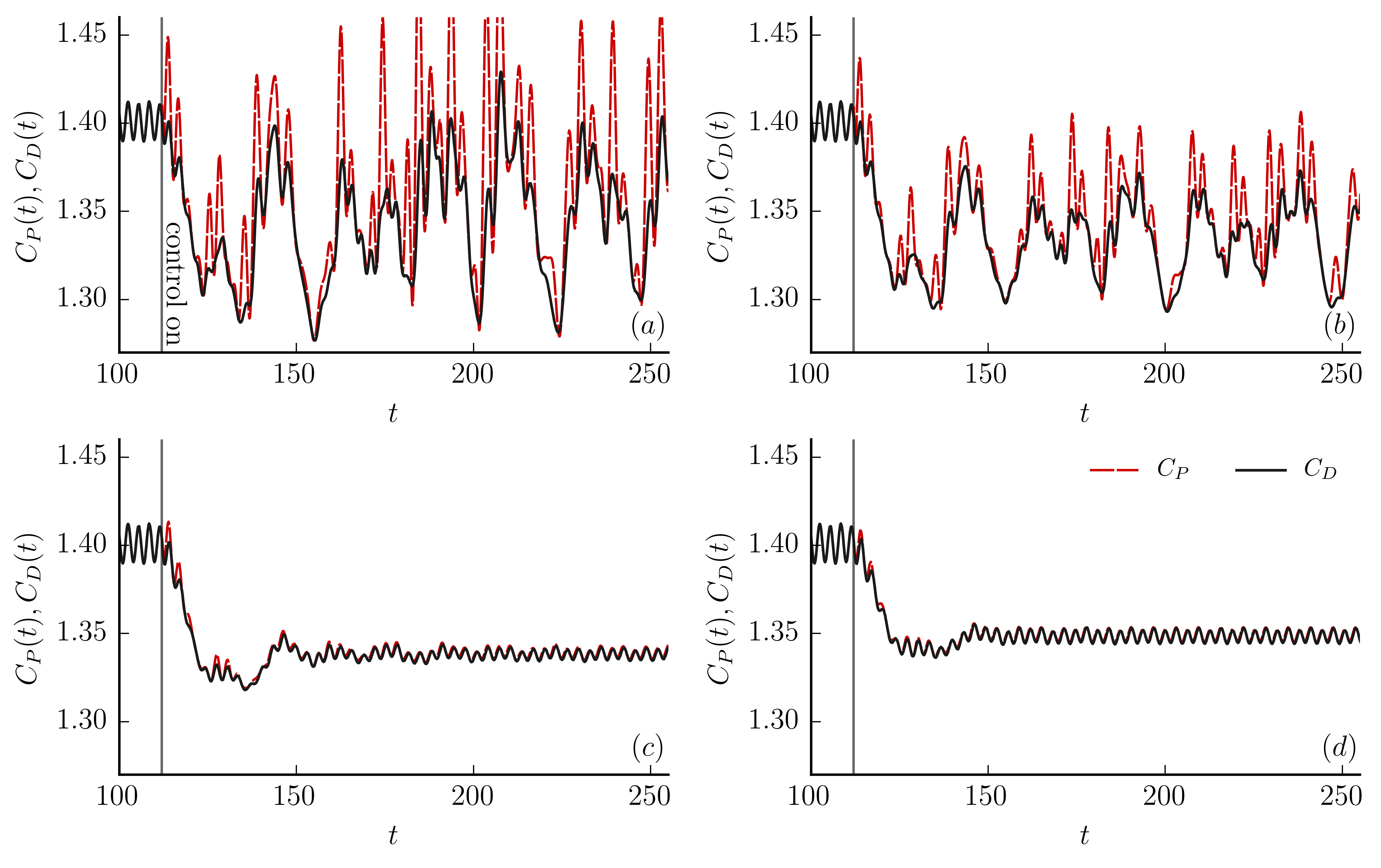}
	\caption{(Colour online). Time histories of the total power and drag coefficients, $C_P(t)$ and $C_D(t)$, obtained from closed-loop direct numerical simulations for the four controllers, with $R=50, 100, 150$ and 200, in panels $(a)$, $(b)$, $(c)$ and $(d)$, respectively. The difference between the two is the energy per unit time transferred to the fluid by the control, i.e. the power spent for actuation.}
	\label{fig:total-power}
\end{figure}

Time histories of the total power coefficient obtained from direct numerical simulation of the closed-loop system, are reported in the four panels of figure \ref{fig:total-power}, for the four controllers derived, with red dashed lines. The drag coefficient is also reported as a black solid line, for reference. The difference between the two, i.e. $C_P - C_D$, is the normalised energy per unit time and unit span required to actively control the flow.

For $R=50$ and $R=100$, panels $(a)$, $(b)$, the drag exhibits a low-frequency modulation similarly to the total fluctuation kinetic energy, with the ``valleys'' of these two quantities matching fairly well. The drag minima can be as low as 1.28, suggesting that the control design methodology is indeed effective, although performance is periodically lost as discussed above. Interestingly, the drag coefficient associated with the steady laminar solution $\mathbf{u}_0$ is, in our setup, 1.14. As a result,  the drag coefficient reduction, compared to the drag associated with vortex shedding \citep{protas2002optimal} can be, instantaneously, as large as 46\%.  The time-averaged drag is of course higher, as reported in table \ref{tab:table-results-dns}. A maximum percentage drag reduction, normalised with the drag coefficient of the uncontrolled flow, of 4.6\% has been obtained, at $R=100$. This value is not as high as in previous closed-loop control studies on this same configuration. Using optimal control theory \citet{protas2002optimal} and more recently \citet{Flinois2015} have achieved drag reductions of 7\% at $Re=75$ and 15\% at $Re=150$, and 19\% at $Re=100$, respectively. However, in these two works, Navier-Stokes equations were used directly for control design, and not a reduction thereof, enabling an effective control strategy to be found. We believe that developing controllers on larger and more accurate ROMs, that correctly describe the change in dynamics as control is activated, will result in increased performance.


In some occasions, the total power coefficient is lower than the drag itself, because the product $C_M(t)\gamma(t)$ is negative. These events indicate that the cylinder is being driven by the viscous torque, corresponding to a passive mechanism where the flow exerts a net work on the cylinder. Nevertheless, this product is positive for most of the time, and indicates that the control is actively manipulating the flow.

For the two larger penalisation factors the long-term cost of the control is extremely small, with peaks of $C_P - C_D$ not exceeding 0.002, practically invisible in the two lower panels of figure \ref{fig:total-power}. The control strategy identified is quite efficient, because in the long-term, a small amount of power is actively spent to reduce the total power by a significantly larger amount. Following \citet{protas2002optimal,bergmann2005optimal}, and based on the definition (\ref{eq:budget}), the efficiency can be quantified by the power saving ratio (PSR)
\begin{equation}
	PSR = \frac{\overline{C_P}^{u} - \overline{C_P}^{c}}{2\overline{C_M\gamma}}
\end{equation}
i.e. the ratio between the power saved and the power spent for control, in a time-averaged sense, where the superscripts $u$ and $c$ indicated the uncontrolled and controlled cases, respectively. The PSR, reported in table \ref{tab:table-results}, is remarkably large for the two higher penalisation factors, when the feedback control operates close to the design point it was constructed for.  For comparison, \citet{protas2002optimal} obtained a PSR equal to 51 at $Re=150$, and 122 at $Re=75$, using optimal control theory in a predictive setting. Various other open-loop-type control approaches, where the cylinder is oscillated harmonically at an optimal frequency and amplitude, are significantly less efficient \citep{bergmann2005optimal}. This is due to the fact that in the present case the feedback controller trades inexpensive control power, scaling directly with the square of the control amplitude and inversely to the square root of the Reynolds number \citep{Bergmann2006ka}, with precious propulsion power mainly associated with the pressure drag.

\section{Discussion and conclusions}\label{sec:conclusions}
The main contribution of the present paper is the development of a novel feedback control design paradigm for incompressible fluid flows. It applies to finite-dimensional dynamical systems given as a set of first-order nonlinear ordinary differential equations, with the right-hand side being a polynomial function in the state variables and in the controls. Galerkin-type models of incompressible fluid flows, obtained from projection of the governing equations on a finite low-dimensional subspace, have exactly this form.

This paradigm of control is rooted on recent advances in control theory and optimisation over polynomials, commonly known as Sum-of-Squares methods. At the core, these methods leverage computationally efficient approaches to construct positive polynomial functions, by formulating and solving convex semidefinite programmes.

The key distinguishing features are that i) the long-term behaviour of the system, the permanent regime, is central in the design stage, i.e. long-time averages of fluctuating quantities can be optimised by control design, and that ii) the nonlinearity is taken directly into account in the design process. Furthermore, the present SOS-based scheme allows the design of polynomial-type feedback controllers of arbitrary degree, hence it is not limited to the linear case discussed here. Further research is required to understand if nonlinear feedback control can considerably improve performance.

We have numerically investigated the problem of mitigating the kinetic energy of velocity fluctuations in the unsteady wake of a circular cylinder at $\Rey=100$, in the laminar regime, via controlled rotary motions of the surface, in a full-information state feedback arrangement. A ten-mode POD-Galerkin reduced-order model of the actuated wake flow was derived. A crucial element is that the phase space of the ROM should host an attractor whose structure is as similar as possible to that of the full-order system when projected on the low-dimensional subspace. This necessity arises from the fact that bound estimation and optimisation via control design target explicitly the attractor of the system and control performance likely increases if the long-term behaviour of the reduced- and full-order system are similar. From this perspective, model calibration schemes that ensure long term stability of the reduced system, and similarity of attractors, \textcolor{black}{at least in a statistical sense}, are desirable, \textcolor{black}{(see e.g. \citet{osth} and references therein).}



Linear state-feedback controllers were derived using the ROM, using a penalisation on the control as a design parameter, and were implemented in direct numerical simulation. These controllers effectively decreased the ``size'' of the limit cycle associated with vortex shedding and significantly reduced the long-time average of the total fluctuation kinetic energy, as well as the time-averaged drag coefficient. The feedback system was energetically efficient, as the power saved per unit control power spent was in the range of 75-80. For lower values of the penalisation factor, the greater control input resulted in better performance just after activation of control, but eventually performance worsened significantly. This is not a limitation of the present SOS-based scheme, but is it rather driven by the POD-Galerkin modelling strategy used, which is known to lack robustness. We expect that improvements in the modelling strategy will result in increased performance in direct numerical simulation.

Currently, the main drawback of this methodology is the unfavourable scaling of the computational cost with the size of the system $N$, and the degree of the polynomial function $V$. Limitations of existing computational tools to pre-process the polynomial inequalities and solve the associated semidefinite programs currently limit the methodology to dynamical systems of size not greater than about 10-20, and the size reduces considerably if high degree polynomials $V$ are used. However, these methods are rather novel and the available computational software tools are designed for generality. Hence, a large number of optimisations can be introduced by specializing these tools to the peculiarities of hydrodynamic-type systems. A few illustrative instances, towards which future efforts will be devoted, are the exploitation of more efficient SDP solvers, sparsity patterns in the right-hand side of the dynamical system as well as in the structure of the tunable function $V$.
  
\textcolor{black}{It is worth to point out that the improvement in our ability to solve the relevant SDP problems necessary for achieving better results might actually be far less than it might seem. In the case of global stability analysis the scalability issue was successfully dealt with in \citet{Huang2015} using the uncertain system method proposed in \citet{Goulart2012692}. This approach can be extended to certain problems of flow control. It allows to construct storage functionals for averaged parameters of systems governed by partial differential equations, that is systems with infinite number of degrees of freedom, while solving the SDP problems corresponding to only a limited number of  degrees of freedom.   The method can also be used to reduce the effective number of degrees of freedom in a finite-dimensional dynamical system. Further details can be found in the cited papers. Here, we only notice that when the bounds are constructed for an uncertain system the corresponding SDP problems have to deal with twice as many independent variables as in the case of a standard (certain in our terminology) system. Hence, the way forward is to construct a better ROM with, for example, 40 modes, reduce it to an uncertain system with, for example, 15 modes, and then build a controller solving an SDP corresponding to 30 independent variables. The required increase in the quality of the SDP solver then corresponds to only 3 times increase in the number of independent variables as compared to the case considered in the present paper. This might indeed become possible in the foreseeable future. }

A second limitation of the methodology is that the controller is formally guaranteed to reduce only the upper bound of a long-time averaged cost function. In a particular realisation of the controlled flow, the actual time-average might not decrease. The essential motivation, discussed at length in the paper, is that control design targets the attractor for which the time-averaged cost is the largest, i.e. the one that the upper bound is associated with. If the system has, for instance, two different attractors, with separate basins of attraction, the control of a trajectory started from an initial condition not in the basin of the attractor associated with the bound, might result in worsened performance. From this perspective, the controller is guaranteed to reduce the time-average only in the worst-case scenario, which might be different from the most likely scenario. In practice, this limitation might be less important than it appears here, although it represents a possibility in the general case.

A further observation is that in this paper we have investigated the control of a dynamical system for which the permanent regime is given by a trivial attractor, i.e. a stable periodic orbit. However, real turbulent flows usually have extremely complicated attractors, and the long-term behaviour is usually chaotic. Reduced-order modelling and long-time average cost control of such type of systems via SOS optimization would be more interesting, but also induce more difficulties. We would like to address them in our future work.

\begin{acknowledgements}
This work was supported by the UK Engineering and Physical Sciences Research Council grants EP/J011126/1, EP/J010537/1 and EP/J010073/1 and received support-in-kind from Airbus Operation Ltd, ETH Zurich (Automatic Control Laboratory), University of Michigan (Department of Mathematics) and University of California, Santa Barbara (Department of Mechanical Engineering). \textcolor{black}{D.H. was partially supported by National
Natural Science Foundation of China under grant No. 61433011}. All data supporting this study are openly available from the University of Southampton repository at http://dx.doi.org/10.5258/SOTON/398622.
\end{acknowledgements}

\bibliography{biblio}
\bibliographystyle{jfm}

\appendix
\section{}\label{sec:trapping region}
A globally attracting absorbing set $\mathcal{T}$ is a closed subset of $\mathbb{R}^N$ for which if $\mathbf{x}(s) \in \mathcal{T}$ it follows that $\mathbf{x}(t) \in \mathcal{T}$ for $t > s$ \citep{Temam2009}. The existence of such a set is a sufficient condition for the trajectories to remain bounded, as for any $\mathbf{x}(0)$ it follows that $\mathbf{x}(t) \in \mathcal{T}$ for all sufficiently large $t$. For dissipative systems, such as fluid flows, the existence of such a set follows from physical considerations. For truncated Galerkin models this can be proven by the search algorithm proposed by \citet{schlegel2015long}. Here, a variant of their approach based on SOS is proposed. We check whether there exists a ball $\mathcal{D}=\{\mathbf{x}\,|\,1/2\mathbf{x}^T\mathbf{x}\le \beta\}$, for a finite positive $\beta$ and containing $\mathcal{T}$, outside of which the system's energy $K=1/2\mathbf{x}^T\mathbf{x}$ is a Lyapunov function, i.e. its time rate of change is \textcolor{black}{negative-definite}:
\begin{equation}
\frac{\mathrm{d}K}{\mathrm{d}t} = \mathbf{x}^T\frac{\mathrm{d}\mathbf{x}}{\mathrm{d}t}=\mathbf{x}^T\mathbf{f}(\mathbf{x}) \textcolor{black}{<} 0 ~~ \forall \mathbf{x} \nsubseteq \mathcal{D}
\label{SOS1}
\end{equation}

Restriction of this polynomial inequality outside of $\mathcal{D}$ can be enforced with application of the $S$-procedure \citep{Tan2006} by introducing and finding a tunable polynomial $S(\mathbf{x})$ satisfying $S(\mathbf{x})\ge 0 ~\forall \mathbf{x}\in{\mathbb R}^N$ for which
\begin{equation}
\left\{
\begin{array}{c}
-\mathbf{x}^T\mathbf{f}(\mathbf{x}) \textcolor{black}{-} S(\mathbf{x})(1/2\mathbf{x}^T\mathbf{x}-\beta) \mbox{~is ~SOS},  \\
[1ex]
S(\mathbf{x})\mbox{~is ~SOS}.
\end{array}
\right.
\label{SOS22}
\end{equation}
Feasibility of this problem for any finite $\beta$ proves the existence of an absorbing set. If problem (\ref{SOS22}) is solved for the minimum $\beta$, the radius of this set can also be estimated. Note that minimising $\beta$  is a convex optimisation problem, so the solution, if it exists, is unique.
It is worth noticing that the infeasibility of the problem cannot disprove the existence of the absorbing set owing to the fact that the positivity constraint has been replaced by an SOS constraint.


\section{}\label{sec:iterative-algorithm}

Recall the reduced-order model of the cylinder flow
\begin{eqnarray}
\displaystyle \frac{\mathrm{d}a_i}{\mathrm{d}t}&=& c_i + {\sum_{j=1}^N L_{ij} a_j}  + {\sum_{j=1}^N \sum_{k=1}^N Q_{ijk}a_j a_k} + {m_i   \frac{\mathrm{d}\gamma}{\mathrm{d}t}  + b_i \gamma^2 + e_i \gamma + \sum_{j=1}^N F_{ij} a_j \gamma} \quad i=1, \ldots N  \nonumber\\
&:=& f_i(\mathbf{a},\gamma,\frac{\mathrm{d}\gamma}{\mathrm{d}t}), \label{eq:ode-sys1}
\end{eqnarray}
where the quadratic term in $\gamma$ vanishes due to the radial symmetry of the control function $\mathbf{u}_c$, as discussed in appendix~\ref{sec:app-ode-sys}.
Define the cost function $\Phi(\mathbf{a})=\Phi_0(\mathbf{a})+ R \gamma^2( \mathbf{a}(t))$, where $\Phi_0(\mathbf{a})=\frac{1}{2}\mathbf{a}(t)^T \mathbf{a}(t) $ and $\gamma( \mathbf{a}(t))$ is assumed to be linear in $\mathbf{a}$ with all the coefficients being undetermined decision variables.
Now, the non-convex SOS optimization problem (\ref{eq:mini-control-sos}) is
\begin{equation}
\begin{array}{c}
~~~~~~~~~~~\underset{V, \gamma}{\text{min}}   ~~~ C \\
[1ex]
\text{subject to}       ~~~ -\big ( \nabla_{\mathbf{a}} V(\mathbf{a}) \cdot \mathbf{f}(\mathbf{a}, \gamma(\mathbf{a}), \frac{\mathrm{d}\gamma(\mathbf{a})}{\mathrm{d}t}) + \Phi_0(\mathbf{a})+R \gamma^2( \mathbf{a}) - C \big ) \in \Sigma.
\end{array}
\label{eq:mini-control-sos1}
\end{equation}
The iterative design algorithm used to solve (\ref{eq:mini-control-sos1}) is given in Table \ref{algorithm}, where equation (\ref{SOS2}) is used.

\begin{table}
  \begin{center}
\def~{\hphantom{0}}
  \begin{tabular}{rl}
      1:  & initial setting $C_0=C_{SOS}^0, \gamma_0=0, I_1=1, I_2=1$ \\
      2:  & \\
      3:  & {\bf while} $I_1=1$ and $I_2=1$ {\bf do} \\
      4:  & ~~~~$d\gamma_0/dt \leftarrow$ calculate the derivative of $\gamma_0$ by (\ref{eq:trick})  \\
      5:  & ~~~~$V\leftarrow$ construct a polynomial with variable coefficients \\
      6:  & ~~~~$expr_1\leftarrow$ construct the expression $-\big( \nabla_{\mathbf{a}} V(\mathbf{a}) \cdot \mathbf{f}(\mathbf{a}, \gamma_0(\mathbf{a}), \frac{\mathrm{d}\gamma_0(\mathbf{a})}{\mathrm{d}t}) + \Phi_0(\mathbf{a})+R \gamma_0^2( \mathbf{a}) - C_0\big )$ \\
      7:  & ~~{\bf if} find a suitable $V$ and an SOS decomposition $expr_1=\mathbf{v}(\mathbf{a})^T{\mathbf Q}_1\mathbf{v}(\mathbf{a})$ {\bf then} \\
      8:  & ~~~~~~$V, {\mathbf Q}_1\leftarrow$ round the coefficients of $V$ and the entries of ${\mathbf Q}_1$ to $d$ decimal places \\
      9:  & ~~~~~~check (\ref{post-check}) for the truncated $V, {\mathbf Q}_1$ using rigorous numerics \\
      10: & ~~~~{\bf if} checks are verified {\bf then} \\
      11: & ~~~~~~~~$I_1=1, C=C_0-\delta C, V_0=V, C_{SOS}=C_0, \gamma_{SOS}=\gamma_0$ \\
      12: & ~~~~~~~~$\gamma\leftarrow$ construct a polynomial controller with variable coefficients \\
      13: & ~~~~~~~~$expr_2\leftarrow$ construct the expression as in (\ref{SOS2}) \\
      14: & ~~~~~~~{\bf if} find a suitable $\gamma$ and an SOS decomposition $expr_2=\mathbf{v}(\mathbf{a},\mathbf{z})^T{\mathbf Q}_2\mathbf{v}(\mathbf{a},\mathbf{z})$ {\bf then} \\
      15:  & ~~~~~~~~~$\gamma, {\mathbf Q}_2\leftarrow$ round the coefficients of $\gamma$ and the entries of ${\mathbf Q}_2$ to $d$ decimal places \\
      16:  & ~~~~~~~~~check (\ref{post-check}) for the truncated $\gamma, {\mathbf Q}_2$ using rigorous numerics \\
      17:  & ~~~~~~~~~~~{\bf if} checks are verified {\bf then} \\
      18:  & ~~~~~~~~~~~~~~$I_2=1, \gamma_0=\gamma, C_0=C$ \\
      19:  & ~~~~~~~~~~~{\bf else} \\
      20:  & ~~~~~~~~~~~~~~$I_2=0$ \\
      21:  & ~~~~~~~~~~~{\bf end} \\
      22:  & ~~~~~~~{\bf else} \\
      23:  & ~~~~~~~~~$I_2=0$ \\
      24:  & ~~~~~~~{\bf end} \\
      25:  & ~~~~{\bf else} \\
      26: & ~~~~~~~~$I_1=0$ \\
      27:  & ~~~~{\bf end} \\
      28:  & ~~{\bf else}  \\
      29:  & ~~~~~~ $I_1=0$ \\
      30:  & ~~{\bf end}  \\
      31:  & {\bf end} \\
      32:  & \\
      33: & {\bf output} $C_{SOS}, \gamma_{SOS}$
  \end{tabular}
\caption{The iterative algorithm used for solving (\ref{eq:mini-control-sos}) with the given ROM (\ref{eq:ode-sys})}
  \label{algorithm}
  \end{center}
\end{table}

\begin{equation}
\mathbf{z}^T\left[
\begin{array}{cc}
-\big( \nabla_{\mathbf{a}} V_0(\mathbf{a}) \cdot \mathbf{f}(\mathbf{a}, \gamma(\mathbf{a}), \frac{\mathrm{d}\gamma_0(\mathbf{a})}{\mathrm{d}t}) + \Phi_0(\mathbf{a})- C\big ) & \gamma(\mathbf{a}) \\
\gamma(\mathbf{a}) & R
\end{array}
\right]\mathbf{z}, ~~\forall \mathbf{z}\in {\mathbb R}^2
\label{SOS2}
\end{equation}

\section{}\label{sec:app-ode-sys}
With $\omega_i(\mathbf{x})$ being the scalar vorticity field associated with the mode $\mathbf{u}_i(\mathbf{x})$, and similarly for $\overline{\mathbf{u}}$ and $\mathbf{u}_c(\mathbf{x})$,  Galerkin projection results in the following coefficients:
\begin{equation}
	c_i = -\frac{1}{Re}\int_\Omega \omega_i \nabla^2 \overline{\omega} \mathrm{d}\Omega - \int_\Omega \mathbf{u}_i \cdot (\overline{\mathbf{u}} \cdot \nabla \overline{\mathbf{u}})\mathrm{d}\Omega,
\end{equation}
\begin{equation}
	L_{ij} = -\frac{1}{Re}\int_\Omega \omega_i \nabla^2 \omega_j \mathrm{d}\Omega - \int_\Omega \mathbf{u}_i \cdot (\overline{\mathbf{u}} \cdot \nabla \mathbf{u}_j)\mathrm{d}\Omega - \int_\Omega \mathbf{u}_i \cdot (\mathbf{u}_j \cdot \nabla \overline{\mathbf{u}})\mathrm{d}\Omega,
\end{equation}
\begin{equation}
    Q_{ijk} = -\int_\Omega \mathbf{u}_i \cdot (\mathbf{u}_j \cdot \nabla) \mathbf{u}_k \mathrm{d}\Omega,
\end{equation}
\begin{equation}
	m_i = - \int_\Omega \mathbf{u}_i \cdot \mathbf{u}_c \mathrm{d}\Omega
\end{equation}
\begin{equation}
	e_i = - \int_\Omega \mathbf{u}_i \cdot (\mathbf{u}_c \cdot \nabla \overline{\mathbf{u}} + \overline{\mathbf{u}} \cdot \nabla \mathbf{u}_c ) \mathrm{d}\Omega - \frac{1}{Re} \int_\Omega \omega_i\omega_c  \mathrm{d}\Omega
\end{equation}
\begin{equation}
	b_i = -\int_\Omega \mathbf{u}_i \cdot (\mathbf{u}_c \cdot \nabla \mathbf{u}_c) \mathrm{d}\Omega
\end{equation}
\begin{equation}
	F_{ij} = -\int_\Omega \mathbf{u}_i \cdot (\mathbf{u}_j \cdot \nabla \mathbf{u}_c + \mathbf{u}_c \cdot \nabla \mathbf{u}_j) \mathrm{d}\Omega
\end{equation}
In the present case all the coefficients $b_i$ are identically zero because of the radial symmetry of the control function $\mathbf{u}_c$. Domain integrals are evaluated numerically on the triangular unstructured mesh by using a linear approximation of the integrand function based on nodal values. All derivatives, for gradients and vorticities, are computed using a local quadratic interpolation scheme available in algorithm 624 from \citet{Renka}. Strictly, some of the above definitions do not contain the line integrals on the boundary of the domain arising from the use of vector calculus identities to eliminate the Laplacian, as in appendix 2 of \citet{bergmann2005optimal}, as these are found to be quite small and negligible in the present case with respect to the domain integrals above. Appropriate symmetries in the tensor $Q_{ijk}$ are numerically enforced after the computations of the integrals to
 impose ensure that the nonlinear term is energy preserving (see e.g. \citet{schlegel2015long} for a discussion on this topic for the present case).

\section{}\label{sec:nonlinear-controllers}

For the ROM (\ref{eq:ode-sys1}), or its compact form
\begin{equation}
\frac{\mathrm{d}{\mathbf a}}{\mathrm{d}t}={\mathbf f}({\mathbf a},\gamma,\frac{\mathrm{d}\gamma}{\mathrm{d}t}),
\label{compact_ode}
\end{equation}
if nonlinear polynomial state-feedback controller is considered, the iterative algorithm shown in Table~\ref{algorithm} cannot be applied directly.
This is due to that the derivative of $\gamma$, as given in (\ref{eq:trick}), is not polynomial any more.
The difficulty can be overcome by regarding $\mathrm{d}\gamma/\mathrm{d}t$ as the virtual control input $u$ and setting a new system state $\tilde{\mathbf a}:=[{\mathbf a}^T \gamma]^T$. The new ROM is
\begin{equation}
\frac{\mathrm{d}\tilde{\mathbf a}}{\mathrm{d}t}={\mathbf f}_1(\tilde{\mathbf a},u),
\label{compact_ode1}
\end{equation}
where the first $N$ equations are same as in (\ref{compact_ode}) while the last one is $\mathrm{d}\gamma/\mathrm{d}t=u$.
As such, the proposed iterative algorithm becomes applicable with minor revision.

\end{document}